\documentclass[twocolumn]{aastex63}
\usepackage[colorinlistoftodos]{todonotes}
\usepackage{amsmath}
\newcommand{\be}{\begin{eqnarray}}
\newcommand{\ee}{\end{eqnarray}}
\usepackage{CJKutf8}
\usepackage{hyperref}
\usepackage{xcolor}

\begin{document}
	
\title{WOCS 4540: Detailed Analysis of a Very Long Orbital Period Blue Straggler}
\begin{CJK*}{UTF8}{gbsn}
\author[0000-0001-9037-6180]{Meng Sun(孙萌)}
\affiliation{Department of Astronomy, University of Wisconsin - Madison, 475 N Charter St, Madison, WI 53706, USA}
\affiliation{Center for Interdisciplinary Exploration and Research in Astrophysics (CIERA), Northwestern University, 1800 Sherman Ave, Evanston, IL 60201, USA}
\author[0000-0002-7130-2757]{Robert D. Mathieu}
\affiliation{Department of Astronomy, University of Wisconsin - Madison, 475 N Charter St, Madison, WI 53706, USA}

\begin{abstract}

WOCS 4540 is the longest orbital period ($P_{\rm orb}=3030$ days) blue straggler (BSS) - white dwarf (WD) pair in the old open cluster NGC 188 . It also contains one of the most luminous BSS in the cluster. Prior \textit{Hubble Space Telescope} COS spectroscopy measured a WD mass of 0.53 $M_{\odot}$, indicative of a carbon-oxygen WD and suggesting previous mass transfer from an asymptotic branch giant (AGB). Detailed modeling of the system evolution, including red giant branch phase wind mass transfer, AGB wind Roche-lobe overflow and regular Roche-lobe overflow, is done with Modules for Experiments in Stellar Astrophysics. The best-fit model produces excellent agreement with a wide array of observational constraints on the BSS, the WD and the binary system.
To produce the observed luminosity and effective temperature of the BSS, all three donor mass-transfer mechanisms contribute similarly to build a 1.5 $M_{\odot}$ BSS. The overall mass-transfer efficiency is 55\%. Regular Roche-lobe overflow occurs only during the largest AGB thermal pulse, but yields a very high accretion rate at 75\% efficiency and briefly (less than 1 Myr) a very high luminosity boost from the accretor.

\end{abstract}


\section{Introduction}
\label{sec:intro}
In star clusters and the Galactic field, multiple sub-populations of stars do not follow single-star stellar evolutionary pathways. One of the longest known examples --- blue straggler stars (BSSs)  --- are more luminous or hotter than the main-sequence (MS) turnoff. Currently, three theories are most frequently discussed for the formation of the BSSs: mass transfer (MT) in binary stars \citep{1964MNRAS.128..147M,1971ARA&A...9..183P,2004MNRAS.355.1182C,2006A&A...455..247T,2008MNRAS.387.1416C},  mergers in close binaries or hierarchical systems \citep{2006ApJ...646.1160A,2009Natur.457..288K,2009ApJ...697.1048P,2011MNRAS.410.2370L,2014ApJ...793..137N,2019MNRAS.488..728F} and direct collisions in dense environments such as cluster cores \citep{1976ApL....17...87H,1983Natur.301..587H,1989AJ.....98..217L,1999ApJ...513..428S}. All of these processes increase the mass of a MS progenitor.

BSSs are common in open clusters and recently have been systematically documented in \cite{2021A&A...650A..67R}. \citet{2021ApJ...908..229L} studied the BSS population in 16 old open clusters. In the well-studied open cluster NGC 188, 14 out of 20 BSSs are binaries with orbital periods greater than 100 days \citep{2009Natur.462.1032M,2015ASSL..413...29M}. This suggested MT as a primary channel in the formation and evolution of the NGC 188 BSSs. This was confirmed by \citet{2014ApJ...783L...8G,2015ApJ...814..163G}, who found white dwarfs (WDs) orbiting around seven of the BSSs. WD companions are direct evidence of a post-MT system, where the evolved donor star loses its envelope, leaving behind its core as a WD, and the accretor star gains mass and becomes a BSS. \citet{2015ApJ...814..163G} concluded that two-thirds of the NGC 188 BSSs formed by MT.

Most of the binary BSSs have orbital periods greater than $\approx$ 1000 days. Given MT origins, such long-period BSSs would be post-case-C MT systems, where MT occurs during the donor's asymptotic giant branch (AGB) phase.

During the AGB phase, stars may lose up to half of their mass through stellar winds \citep{2004ApJ...604..800W}. Wind mass-loss rates can reach $10^{-8}$ to $10^{-4}\,M_{\odot}/{\rm yr}$ \citep{2018A&ARv..26....1H}. In addition to MT via Roche-lobe overflow (RLOF), \citet{2007ASPC..372..397M} proposed MT through wind Roche-lobe overflow (WRLOF) in AGB binaries, and modeled the process with hydrodynamic simulations. Later, \citet{2013A&A...552A..26A} fit a function to their WRLOF efficiency to enable implementation in 1-D binary evolution codes. 

BSSs in open clusters are powerful experiments to better understand such processes of MT, and binary evolution theory in general. In addition to the detailed properties of the two stars and the binary orbital parameters, their presence in clusters provides key information such as age, turnoff masses, and metallicity.

This paper provides a detailed study of the long-orbital period BSS-WD binary WOCS 4540 in NGC 188. Both RLOF and WRLOF MT are included in modeling the formation and evolution.

The paper is organized as follows. Section \ref{obs_fact} summarizes the detailed observational data available for WOCS 4540. Section \ref{Detailed Binary Simulations of WOCS 4540} presents the physical implementation of WRLOF. Section \ref{best_model} introduces our best-fit model of WOCS 4540, incorporating both RLOF and WRLOF within MESA, and describes the important processes and changes during the system evolution. Section \ref{sec: RLOF-Only Model} presents a RLOF-only model for comparison. Section \ref{sec:Discussion} discusses the luminosity during the accretion, stellar rotation and the role of forming systems similar to WOCS 4540 via mergers in hierarchical systems. We conclude with the main findings of this paper in Section \ref{conclusion}.

\section{The Observed Properties of the WOCS 4540 System}
\label{obs_fact}

The binary nature of WOCS 4540 was discovered by \citet{2008AJ....135.2264G}, with a spectroscopic orbital solution presented by \citet{2009AJ....137.3743G}. The very long 3030-day period and the presence of a carbon-oxygen (CO) WD companion \citep{2019ApJ...885...45G} indicate an AGB MT origin. 

The orbital eccentricity is 0.36. While such a large eccentricity is no longer a surprise for post-MT systems (e.g., \citealt{2011Natur.478..356G,2015ASSL..413..153B}), a full explanation for the survival or generation of such eccentricity during the MT process is needed.

WOCS 4540 is among the most luminous of the 20 known BSS in NGC 188. The BSS effective temperature is $6590 \pm 100~ {\rm K}$ \citep{2015ApJ...814..163G}. WOCS 4540 sits well to the red of the ZAMS, but still to the blue of the evolved MS (see Figure 5 of \citealt{2015ApJ...814..163G}). 

\citet{2015ApJ...814..163G} found WOCS 4540 to have the hottest WD companion among the NGC 188 BSSs, with a photometric temperature of $18100\pm 500$ K and an inferred cooling age of only 70 Myr. 

More recently, \citet{2019ApJ...885...45G} undertook a detailed analysis of a far-ultraviolet COS spectrum (central wavelength 1105 $\mathrm{\AA}$) of WOCS 4540. They find the WD companion to have an effective temperature $T_{\rm eff}$ = 17,000 K and a surface gravity $\log\,g$ = 7.80 cm s$^{-2}$. They find a mass for the secondary of $0.53\pm 0.03 M_{\odot}$, which they interpret as a CO WD, and a cooling age of 105 Myr.

Finally, the presence of WOCS 4540 in a well-studied open cluster is of great benefit. \citet{2019ApJ...885...45G} review current distance estimates to NGC 188; following them we adopt a Gaia-based distance of 1845 $\pm$ 107 pc. The uncertainty encompasses other distance estimates. We also adopt $E(B-V)=0.09$ \citep{1999AJ....118.2894S} and solar metallicity \citep{1990AJ....100..710H,2011AJ....142...59J}. Age determinations for NGC 188 vary from 6.2 $\pm$ 0.2 Gyr to 7 $\pm$ 0.5 Gyr \citep{2009AJ....137.5086M,1999AJ....118.2894S}, while the Bayesian analysis of
\citet{2015AJ....149...94H} shows systematic uncertainties in NGC 188 age determinations of 0.7 Gyr. As a reference point, with MESA modeling we find the turnoff mass of NGC 188 to be 1.1 $M_{\odot}$.

Table \ref{observation vs model} lists the observed quantities of the system and of the two stars.

\section{Detailed Binary Simulations of WOCS 4540}
\label{Detailed Binary Simulations of WOCS 4540}

\subsection{Overview}

We follow an approach similar to the detailed MT modeling of the NGC 188 blue straggler WOCS 5379 \citep{2021ApJ...908....7S}. The initially more massive primary star produces the current WD from its core. The initially less massive secondary becomes the BSS. We assume that the two progenitor stars were born in the binary and evolved simultaneously. 

The evolutionary simulation for WOCS 4540 uses the ``evolve\_both\_stars" option in the Modules for Experiments in Stellar Astrophysics code (MESA, version 12115; \citealt{2011ApJS..192....3P,2013ApJS..208....4P,2015ApJS..220...15P,2018ApJS..234...34P,2019ApJS..243...10P}). This standard option in MESA incorporates MT through regular RLOF  (and the Bondi-Hoyle-Lyttleton (BHL; \citealt{1944MNRAS.104..273B}) theory of fast wind accretion, which is turned off for this work).

In addition, the simulations for WOCS 4540 in this paper include MT through WRLOF  (\citealt{2010AIPC.1314...51M,2013A&A...552A..26A,2015ASSL..413..153B}). The implementation of this process in MESA is described in Section \ref{sec:wind MT}.

Non-conservative MT is permitted in our simulations, where a fraction of the mass from the donor star leaves the system.
In this work, we define overall MT efficiency as the mass change of the accretor star divided by the mass change of the donor star. This overall MT efficiency is the result of both RLOF and WRLOF. We note that in MESA, MT efficiency is defined as the percentage of the mass lost by the donor star that is accreted onto the companion through only regular RLOF, which we distinguish as RLOF MT efficiency. In the simulations here, the RLOF MT efficiency is taken to be constant throughout the MT. Finally, WRLOF MT efficiency is described in detail in Section \ref{sec:wind MT}. 

Material lost from the system is taken to escape from the vicinity of the accretor by undefined processes (the $\beta$ mechanism). As discussed in \cite{2021ApJ...908....7S}, this choice, as compared to escape from the vicinity of the donor for example, does influence the evolutionary path of the accretor. However the differences are not so large as to change the ability to reproduce the observed BSS. As such, the observations cannot constrain the loss location.

After the donor and the accretor transfer material via RLOF or WRLOF or both, they adjust to new hydrostatic and thermal equilibria. During such evolution, the accretor radius is required to be smaller than its Roche-lobe radius. If the accretor fills its Roche-lobe radius, the simulation is stopped because a contact binary has been made, which does not agree with the observations of WOCS 4540. In the case of non-conservative MT, system angular momentum is reduced, which mitigates the formation of such contact. 

The inlists which contain general solar-type star settings and the new code implementing WRLOF (Section 3.2) are shared at \url{https://zenodo.org/record/7402330#.Y453qi2z2_V}. Following \cite{2021ApJ...908....7S}, we adopt for solar metallicity $Z=0.019$. Reimers' and Bl\"{o}cker's red giant branch (RGB) and AGB wind schema are employed in the simulation, with scaling factors as given in Table \ref{initial model}.

In this modeling, we do not consider the evolution of the spins of the two stars or the orbital eccentricity. We recognize the importance of both and plan to incorporate both in future work.

\subsection{Wind Roche Lobe Overflow}
\label{sec:wind MT}

The WRLOF mechanism was first introduced by \citet{2007ASPC..372..397M,2010AIPC.1314...51M} to explain observations of the Mira binary, which contains a pulsating AGB star and a compact object (usually assumed to be a WD). {\it Hubble Space Telescope} ultraviolet observations show that there is mass outflow from Mira A to Mira B, but the radius of the donor star is still much smaller than the Roche-lobe radius. Neither BHL accretion nor regular RLOF could explain the observations. More recently, to increase the estimated fraction of carbon-enhanced metal-poor (CEMP) stars in binary population syntheses, \citet{2013A&A...552A..26A} applied WRLOF in a fast binary population synthesis code. The result of the study shows that the disagreement between theory and observation regarding the population of CEMP stars can be reduced with the inclusion of WRLOF.

Physically, WRLOF is conceptualized in the context of slow winds, such as found for example from AGB stars. The effective temperatures of AGB stars can be low enough to form dust grains in the extended atmospheres, especially during thermal pulses. Those dust grains can be accelerated by radiation pressure to drive a slow wind (of both dust and gas) escaping the star. If the wind acceleration radius (where the wind reaches escape velocity) lies near or beyond the Roche lobe, then within the Roche lobe the AGB wind is gravitationally guided through the inner Lagrangian point. In addition, a binary companion in the wind acceleration zone can have an orbital speed comparable to the wind overflowing the Roche lobe, resulting in dynamical focusing of the flow onto the accretor and MT efficiencies as high as 50\%. 

\citet{2013A&A...552A..26A} calculated the WRLOF MT efficiency $\beta_{\rm acc}$ from hydrodynamical simulations, and then fit $\beta_{\rm acc}$ with a simple function of mass ratio, the main chemical composition of the wind, Roche-lobe radius and the radius of the wind acceleration zone. This formalism can be used in 1-D stellar evolutionary codes, such as this work. 

In general, the MT efficiencies of the three mechanisms rank as follows: BHL accretion is the least efficient, WRLOF efficiency is between BHL accretion and regular RLOF, and regular RLOF is the most efficient \citep{2007ASPC..372..397M}.

For WOCS 4540, MT occurs during the donor's AGB phase, where stellar winds are strong and can reach $10^{-7}\,M_{\odot}/{\rm yr}$. Thus, WRLOF is considered in our simulation. Specifically, we have implemented the \citet{2013A&A...552A..26A} description of the wind MT in the MESA code. $\beta_{\rm acc}$ is a function of the Roche-lobe radius and the radius of the wind acceleration zone as below,

\be
\beta_{\rm acc} = c_1 \left( \frac{R_{\rm wind}}{R_{\rm RL}} \right)^2+c_2\left( \frac{R_{\rm wind}}{R_{\rm RL}}\right)+c_3.
\ee
Here $c_1$, $c_2$ and $c_3$ are fitting coefficient; we adopt $c_1 = -0.284$, $c_2 = 0.918$ and $c_3 = - 0.234$ following \citet{2013A&A...552A..26A}. $R_{\rm RL}$ is the Roche-lobe radius of the donor star. $R_{\rm wind}$ is the wind acceleration-zone radius. Outside of this radius, the extended material is driven outward by the stellar wind. To calculate this radius, we use Equation (1) from \citet{2013A&A...552A..26A},

\be
R_{\rm wind} = \frac{1}{2}R_{\rm d}\bigg(\frac{T_{\rm eff}}{T_{\rm cond}}\bigg)^{2.5},
\ee
where $T_{\rm cond}$ is the condensation temperature of the dust, which depends on the composition of the dust. In our case, we applied $T_{\rm cond}=1500\,{\rm K}$ for carbon-rich dust. $R_{\rm d}$ is the radius of the donor star.

We also applied the mass-ratio dependence of \citet{2013A&A...552A..26A} in calculating the WRLOF accretion efficiency, in particular for the large system-separation case (e.g. when the wind speed is comparable to the orbital velocity). The final WRLOF efficiency is 

\be
\beta_{\rm acc} = \mathrm{min} \left\{\frac{25}{9}q^2 \beta_{\rm acc},\beta_{\rm acc,max} \right\}.
\label{wind MT efficiency}
\ee

Again following \citet{2013A&A...552A..26A}, the maximum efficiency $\beta_{\rm acc,max}$ is set at a value of 0.5 as found by numerical simulations; none of the models of \cite{Mohamed2010phd} has an efficiency greater than 0.5.

We note that Equation \ref{wind MT efficiency} was only developed by \citet{2013A&A...552A..26A} for the AGB phase. Thus we apply equation \ref{wind MT efficiency} when the central helium mass fraction drops below 0.0001 (i.e., the end of the horizontal branch evolution). To date, WRLOF during the RGB hasn't been studied systematically (private communication, O. Pols). For RGB wind MT we adopt a constant efficiency of 50\%, the maximum WRLOF efficiency found by \citet{Mohamed2010phd}. While adopting this value for RGB winds is somewhat arbitrary, we also find that our best-fit model (next section) only is able to reproduce the large luminosity (and presumably mass) of WOCS 4540 with such a RGB wind MT efficiency.

\section{The Best-Fit Model for WOCS 4540}

\label{best_model}

This section first shows how we search for the best-fit (BF) model of WOCS 4540, one which matches most closely the observed values of the binary system and both stars (e.g., orbital period, BSS luminosities and effective temperatures, WD $\log\,g$ and $T_{\rm eff}$). Then the next three subsections provide detailed discussions of the BF model.

\subsection{Finding the Best-Fit Model for WOCS 4540}

The MT phase is always much shorter than the MS lifetime of the donor star. Thus the initial mass of the evolved donor star is largely defined by the current age of NGC 188, between 6.2 to 7 Gyr, but is permitted to be somewhat larger than the current turnoff mass. We began our search for the BF model with a 1.2 $M_\odot$ donor.

Using a single-star evolution model for WOCS 4540, \cite{2019ApJ...885...45G} estimated the BSS in WOCS 4540 to be around 1.5 - 1.6 $M_{\odot}$. They measured the companion to be a 0.53 $M_{\odot}$ CO WD. Thus the overall MT efficiencies were searched from 50\% to 100\% and initial accretor masses were searched from 0.9 to 1.15 $M_{\odot}$. 

The presence of a CO WD dwarf indicates MT from an AGB donor. The initial orbital period was searched from 1000 to 1500 days to reach the final observed period of 3030 days. Prior to and during the AGB MT, the orbital period increases due to strong stellar wind mass loss from the donor. This acts to stabilize the onset of MT. The orbit also expands with mass reversal during the MT. 

As the last step, a fine-tuning of the initial accretor and donor masses, the initial orbital period, and the RLOF efficiency was done to best match the observations. The fine-tuning steps for donor mass and accretor mass were 0.01 $M_{\odot}$, for initial orbital period was 10 days, and for RLOF efficiency was 5\%.

We note that the parameter search ranges were the same for this BF model (with RLOF and WRLOF) and for the RLOF-Only model (Section \ref{sec: RLOF-Only Model}).

The metrics for goodness of fit were orbital period, accretor luminosity and effective temperature, WD gravity and effective temperature, and system age (which implicitly incorporates the very short WD cooling time).

Models generated around the BF model vary continuously within the observational uncertainties of these metrics. We have explored the ranges of the key physical parameters of the BF model that vary the observable model outcomes within their uncertainties. We find plus/minus ranges on the donor mass of 0.01 $M_{\odot}$, on the accretor mass of 0.01 $M_{\odot}$, on the initial orbital period of 20 days, and on the RLOF MT efficiency of 15\%.

We have not done a complete exploration of parameter space; this task is beyond the scope of this paper. Rather, as described above, we have used physical arguments to focus the model parameter domain. We note again how comprehensively constrained is the physical state of WOCS 4540 by its observations and by its membership in an open cluster. It is very unlikely that a completely different domain of parameters can satisfy all of the constraints as this model does. Of course, alternative physical frameworks should be investigated.

In the following description, the properties of the donor star (progenitor of the WD) have the subscript ``d''. The properties of the accretor star (progenitor of the BSS) have the subscript ``a''.

\subsection{The Best-Fit Model for WOCS 4540}

Table \ref{initial model} shows the progenitor binary of the BF model, along with physical properties of the MT itself. The system starts with a binary comprising 1.20 $M_{\odot}$ and 1.14 $M_{\odot}$ stars in a 1330-day circular orbit. The RLOF MT efficiency is 75\%, while the final overall MT efficiency is 55\%. The product is a 1.50 $M_{\odot}$ BSS and a 0.55 $M_{\odot}$ CO WD in a 2961-day orbit. 

Table \ref{observation vs model} compares the final binary of the BF model with the observations of WOCS 4540. All of the model results agree well with the observations. This agreement includes the model WD, whereas the WOCS 5379 model WD underestimated the measured mass by 0.1 $M_{\odot}$, larger than the observational uncertainty.

\begin{table*}[]
\caption{Initial binary and MT properties for the Best-Fit model}
\begin{center}
\begin{tabular}{c c}
\hline\hline 
Initial conditions (Unit)	&	value	\\
\hline
$M_{\rm 1,i}$ ($M_{\odot}$) & 1.20\\
$M_{\rm 2,i}$ ($M_{\odot}$) & 1.14\\
$P_{\rm orb,i}$ (days) & 1330\\
RLOF efficiency & 75\% \\
RGB wind MT efficiency & 50\% \\
AGB wind MT efficiency & Equation \ref{wind MT efficiency} \\
Reimers scaling factor & 0.5 \\
Bl\"{o}cker scaling factor & 0.1 \\
\hline
Overall MT efficiency & 55.4\% \\
Donor's mass loss through RGB wind ($M_{\odot}$) & 0.19 \\
Donor's mass loss through AGB wind ($M_{\odot}$) & 0.29\\
Donor's mass loss through regular RLOF ($M_{\odot}$) & 0.15\\
Accretor's mass gain through RGB wind ($M_{\odot}$) & 0.10 \\
Accretor's mass gain through AGB wind ($M_{\odot}$) & 0.15\\
Accretor's mass gain through regular RLOF ($M_{\odot}$) & 0.11\\
\hline
\label{initial model}
\end{tabular}
\end{center}
\end{table*}

\begin{table*}[]
\caption{Properties of WOCS 4540 and the Best-Fit model}
\begin{center}
\begin{tabular}{c c c}
\hline\hline 
Property (Unit)	&	Observation& Best-Fit Model	\\
\hline
BSS $T_{\rm eff}$ (K) & $6590^{+100}_{-100}$ & 6650 \\
BSS Luminosity ($L_{\odot}$) & 	 9.8$\pm$0.9 &  9.49	\\
BSS Mass ($M_{\odot}$) & 	 -  &  1.50	\\
WD $T_{\rm eff}$ (K)& $17000\substack{+140 \\ -200}$& 17060 \\
WD $\log\,(g/(\rm cm\,s^{-1})$) &	$7.80\substack{+0.06 \\ -0.06}$ & 7.88	\\
WD mass ($M_{\odot}$) &	$0.53\substack{+0.03 \\ -0.03}$ & 0.55	\\
$P_{\rm orb}$ (day) & 3030$\pm$70& 2961 \\
Age (Gyr) & 6.2 - 7.0 & 6.4\\
Transformation Age (Myr) &  -  & 120 \\
Metallicity & 0.019 & 0.019 \\
Eccentricity & 0.36 & 0  \\
\hline
\label{observation vs model}
\end{tabular}
\end{center}
\end{table*}

\begin{figure}[tp]
\includegraphics[width=0.5\textwidth]{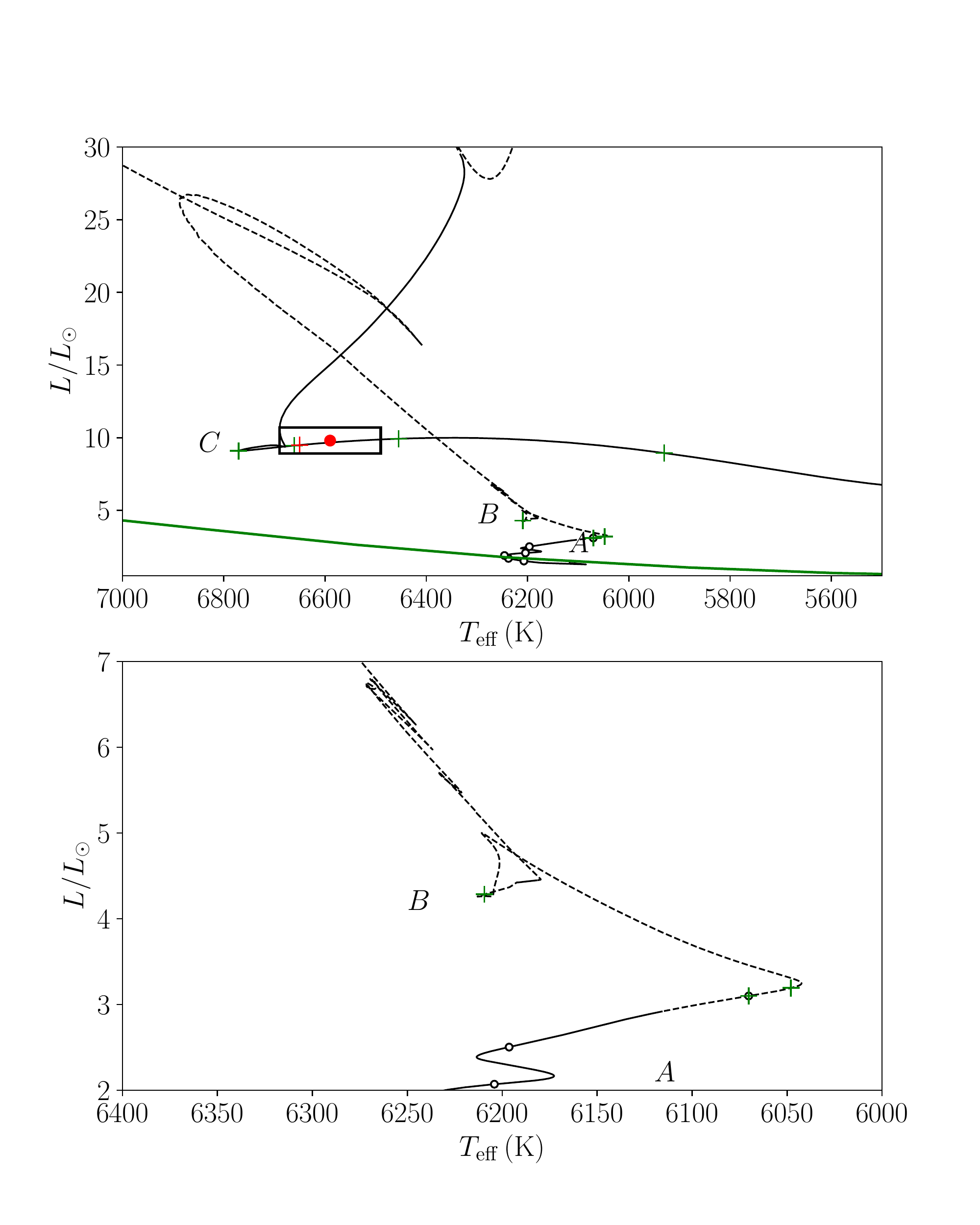}
\caption{Top panel: The HR diagram ($L$ versus $T_{\rm eff}$) for the evolution of the secondary star into a BSS. The open circles are placed at every 1 Gyr. The phases with a total mass accretion rate $<10^{-12}M_{\odot}/\rm yr$ are shown by the solid line. The mass accretion phase is shown by the dashed line. The red dot represents the observed $L$ and $T_{\rm eff}$ for WOCS 4540 BSS. The red dot shows the current state of the BSS, with the surrounding box indicating the measurement uncertainties. The green solid line is the ZAMS line from \citet{2015ApJ...814..163G}. Bottom panel: Zoomed-in around point A and B in the top panel.}
\label{HR2_zoomin.pdf}
\end{figure}

Figure \ref{HR2_zoomin.pdf} presents the BF evolutionary track of the accretor star (the BSS). The MT phase is shown with a dashed curve. The observed bolometric luminosity $L$ and the measured effective temperature $T_{\rm eff}$ \citep{2019ApJ...885...45G} of the BSS are shown with the red dot. The box around the red dot shows the 1-$\sigma$ observational uncertainties.  

The simulation begins with the 1.2 $M_{\odot}$ donor star on the ZAMS near $(T_{\rm eff}/{\rm K},\, L/L_{\odot})=(6225, 1.6)$, and the 1.14 $M_{\odot}$ accretor star on the ZAMS near $(T_{\rm eff}/{\rm K},\, L/L_{\odot})=(6084, 1.28)$, both with small convective cores and convective envelopes near the surface.

The MT begins at position A in Figure \ref{HR2_zoomin.pdf}. The accretor has just evolved off the MS and its central hydrogen is exhausted. At this time the donor is in the RGB phase of evolution. The radius of the donor star has not filled the Roche-lobe radius, so the MT onto the accretor is all from the donor's wind. The bottom panel in Figure \ref{HR2_zoomin.pdf} zooms in on the portion of the top panel near position A and B. The small triangle near B in the bottom panel is when the donor moves to the horizontal branch.

After position B in Figure \ref{HR2_zoomin.pdf}, the donor star enters the phase of AGB thermal pulses. At this time the donor has two thin nuclear burning regions, the helium and hydrogen burning shells. The thin hydrogen burning shell prevents the inner helium shell from burning stably, leading to multiple thermal pulses. The thermal pulse phase is short, of order 0.4 Myr. 

The donor star changes its radius rapidly with these pulses, which leads to significant changes in $\dot{M}$ as well. With the exception of one pulse, the MT during this phase is primarily due to WRLOF, which increases with each increase in donor radius. The evolutionary path of the accretor star moves back and forth, in both $T_{\rm eff}$ and $L$, due to the large variations in the accretion rate driven by these thermal pulses. With the end of the the shell burning of the donor star, the MT ends. 

\begin{figure}[tp]
\includegraphics[width=0.5\textwidth]{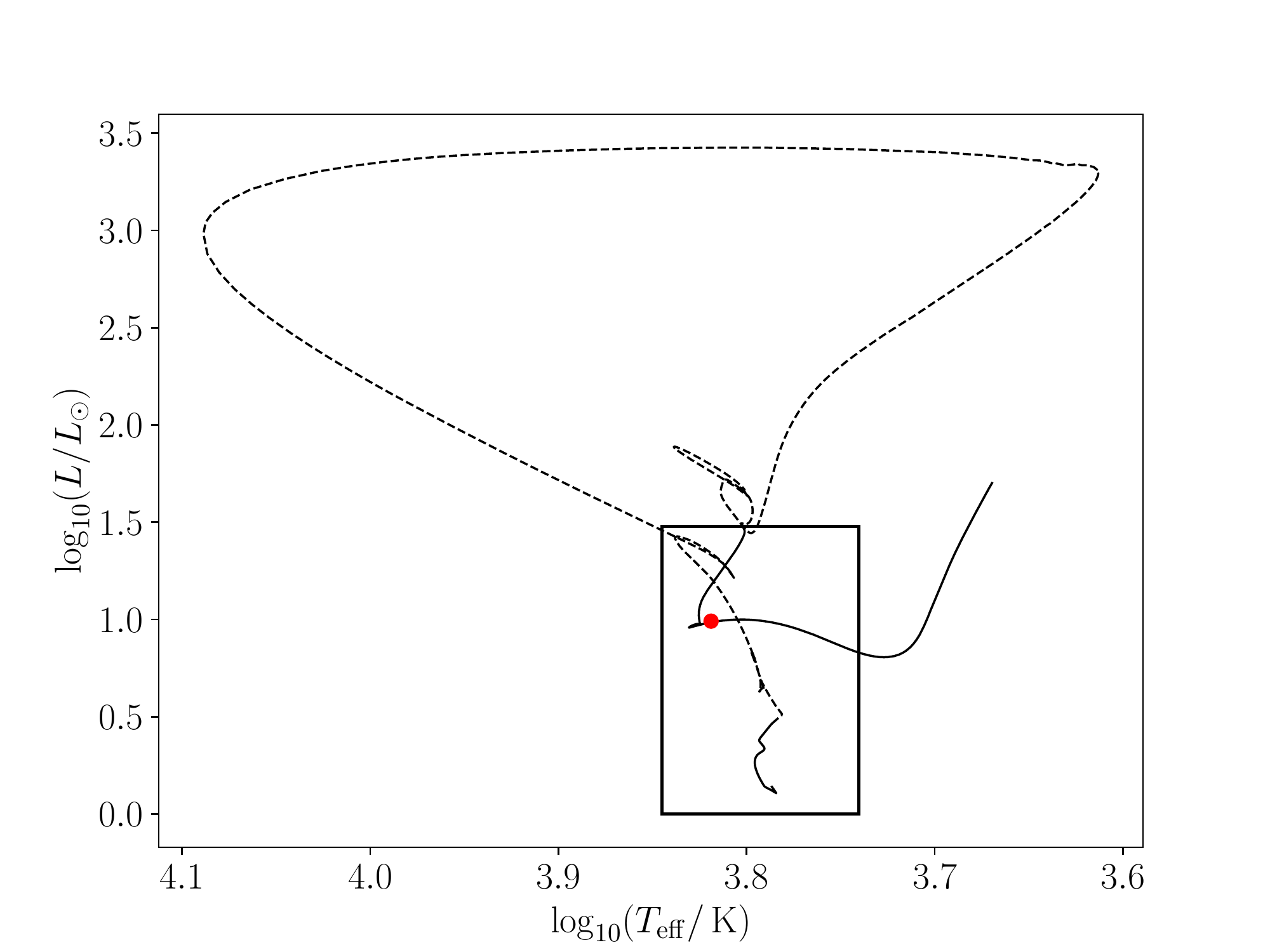}
\caption{The zoom-out version of the Figure \ref{HR2_zoomin.pdf}. The figure description is the same as Figure \ref{HR2_zoomin.pdf}. {The box indicates the same range of the top panel of Figure \ref{HR2_zoomin.pdf}.}}
\label{HR2.pdf}
\end{figure}

During one especially large mass accretion event, the accretor star briefly reaches very large luminosities. Figure \ref{HR2.pdf} is the zoomed-out version of Figure \ref{HR2_zoomin.pdf}. The loop on the evolutionary track occurs via regular RLOF during the largest thermal pulse of the donor star (see detailed discussion in later sections). The accretor star rapidly reaches $\sim 2500$ $L_{\odot}$ on a short timescale of less than 0.1 Myr. At the maximum luminosity, the mass accretion rate is $\sim 10^{-3} M_{\odot}/{\rm yr}$. During the loop, the large luminosities are mainly supported by the gravitational heating of the accreting material.  

After the end of the MT, the accretor star moves to an evolved MS position of a 1.5 $M_{\odot}$ star, burning hydrogen into helium outside the small helium core formed before the MT.

At position C the accretor star transitions from core hydrogen burning to a hydrogen-burning shell, near 6.3 Gyr. The accretor star also begins to develop an outer convective envelope on its way to becoming an RGB star. 
Thus the BSS in WOCS 4540 is a 1.5 $M_{\odot}$ subgiant star at the observed position shown as a red dot. 

In Figure \ref{HR2_zoomin.pdf}, the open circles are placed at every 1 Gyr. The green crosses are placed at every 0.1 Gyr from 6.0 to 6.8 Gyr. 

In summary, the accretor star starts as a 1.14 $M_{\odot}$ ZAMS star with a radiative core and a convective envelope. After exhaustion of H in the core, MT onto the accretor begins through both WRLOF and regular RLOF. The MT rate is variable due to thermal pulses in the AGB donor, resulting in associated variations in the accretor luminosity and effective temperature. During one pulse, the accretor briefly becomes 6000 K hotter and more luminous by three orders of magnitude. When the MT ends, the accretor is a 1.5 $M_{\odot}$ BSS burning hydrogen in a convective core. After the core hydrogen is exhausted, the BSS ignites a hydrogen shell and enters the subgiant branch where is currently resides. At the ascending RGB phase of the accretor star, the simulation is terminated. 

\begin{figure}[tp]
	\includegraphics[width=0.5\textwidth]{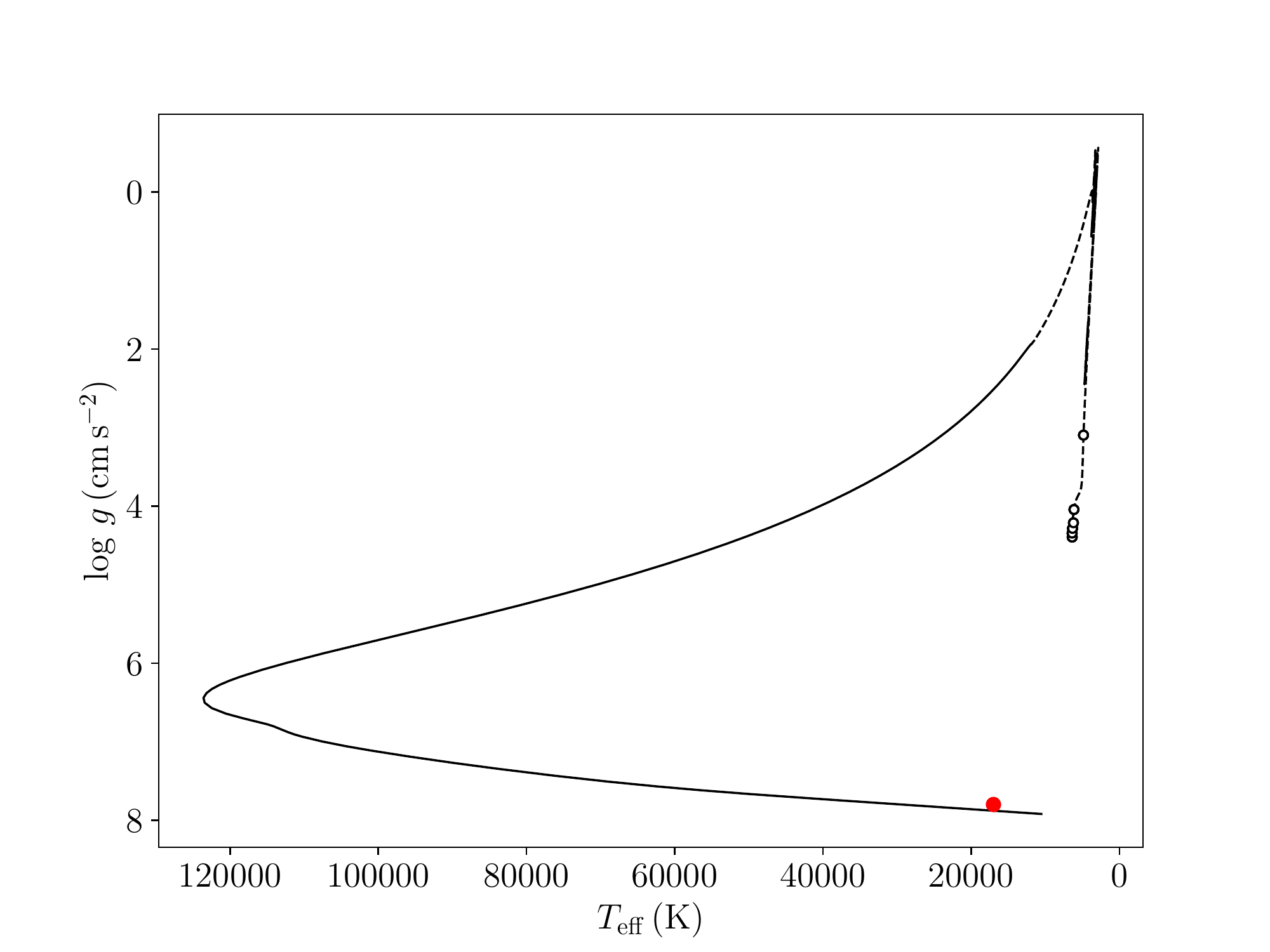}
	\caption{The evolutionary track of the primary star on the $\log g$ - $T_{\rm eff}$ plane. The red dot shows the position of the WOCS 4540 WD. The uncertainty is smaller than the dot size. The dashed line gives the total mass loss rate of the donor star, which is greater than $10^{-12}\,M_{\odot}/{\rm yr}$. The open circles place at every 1 Gyr.}
	\label{HR_WD.pdf}
\end{figure}

\begin{figure}[tp]
	\includegraphics[width=0.5\textwidth]{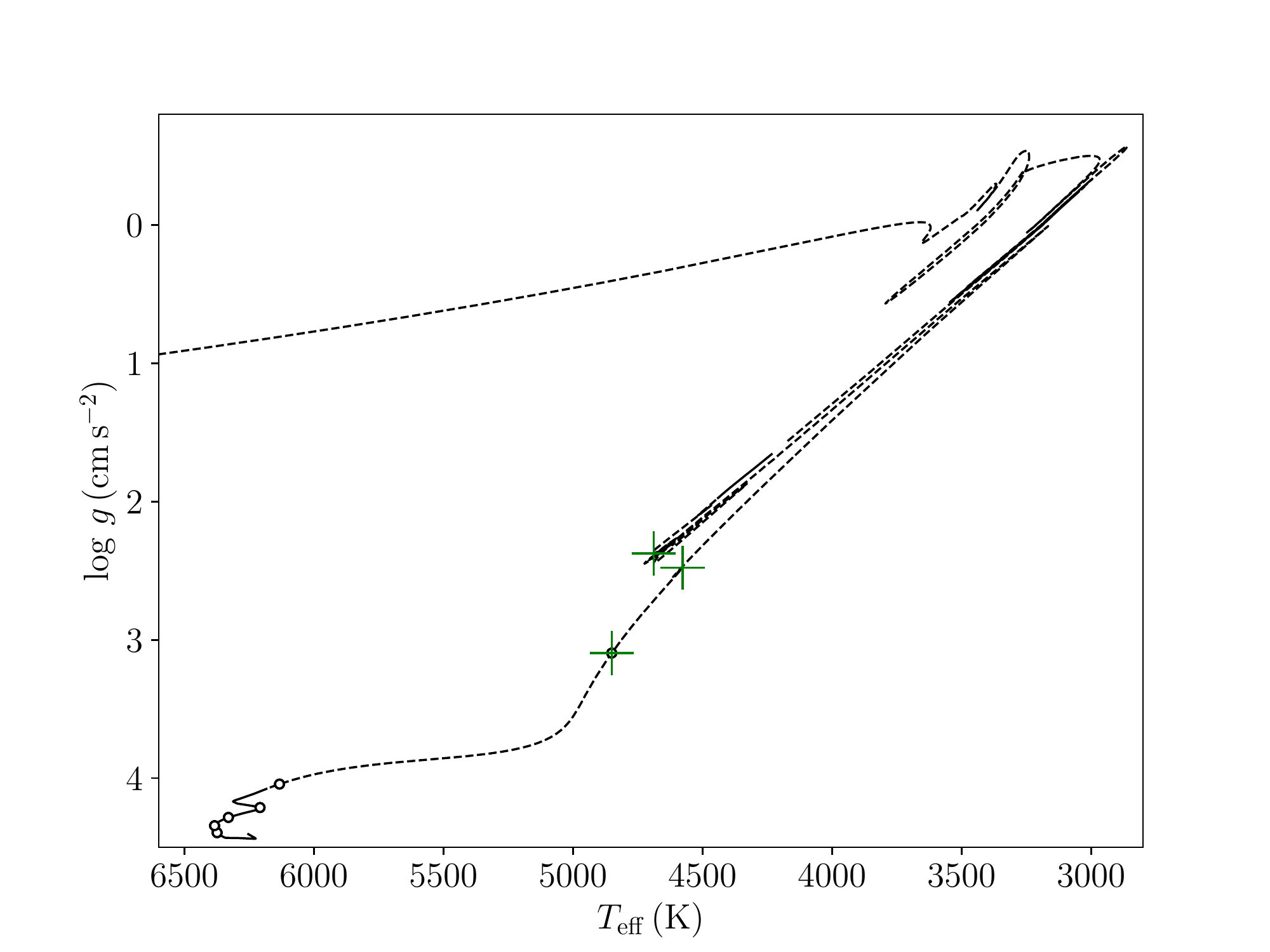}
	\caption{The zoom-in version of Figure \ref{HR_WD.pdf}, emphasizing the domain of the thermal pulses. The open circles are placed at every 1 Gyr. The green crosses are placed at every 0.1 Gyr starting from 6.0 Gyr.}
	\label{HR_WD_zoomin.pdf}
\end{figure}

Figure \ref{HR_WD.pdf} shows the evolutionary track of the donor star, the WD progenitor, in the $\log g$ - $T_{\rm eff}$ plane; Figure \ref{HR_WD_zoomin.pdf} shows a zoomed-in version from the ZAMS through the AGB phase. The evolution starts at $(T_{\rm eff},\,\log g)=(6226,4.44)$. The initial wind mass loss rate for this $1.2M_{\odot}$ MS star is about $10^{-12}$ to $10^{-11}\,M_{\odot}/{\rm yr}$. Less than $0.01 M_{\odot}$ is lost through the MS stellar wind. Near 5.6 Gyr, the donor star is at the base of the RGB. The star spends 0.6 Gyr through the RGB phase, where the wind mass loss rate is between $10^{-11}$ to $10^{-7}\,M_{\odot}/{\rm yr}$.  By the time it reaches the tip of the RGB branch, it has lost 0.19 $M_{\odot}$ in total. At the same time, the wind mass loss expands the binary separation as the system specific angular momentum increases. 

In the late RGB phase, the donor star has developed a 0.46 $M_{\odot}$ helium core. The core helium ignition is at 6.17 Gyr, after which the star enters the horizontal branch phase. The radius of the star shrinks and $T_{\rm eff}$ increases.

With completion of core helium burning, the donor star enters the AGB phase. The thermal pulses start at 6.28 Gyr. In Figure \ref{HR_WD_zoomin.pdf}, the open circles mark every 1 Gyr. The green crosses mark every 0.1 Gyr since 6 Gyr. The dashed line shows the donor's mass loss phase.

During the AGB phase, the donor star loses a significant fraction of its mass, from initially $1\,M_{\odot}$ to $0.56\,M_{\odot}$. Table \ref{initial model} lists the amount of the mass loss at different phases. 

The tip of the AGB is near $(T_{\rm eff},\,\log g)=(2856,-0.56)$. With the end of helium shell burning and a mass of 0.55 $M_{\odot}$, the donor star enters the pre-WD phase. After $(T_{\rm eff},\,\log g)=(123600,6.4)$, the WD is in the cooling phase and passes through the measurements of the observed WD.

\subsection{Mass Transfer in the Best-Fit Model}

\begin{figure*}[tp]
\includegraphics[width=1.0\textwidth]{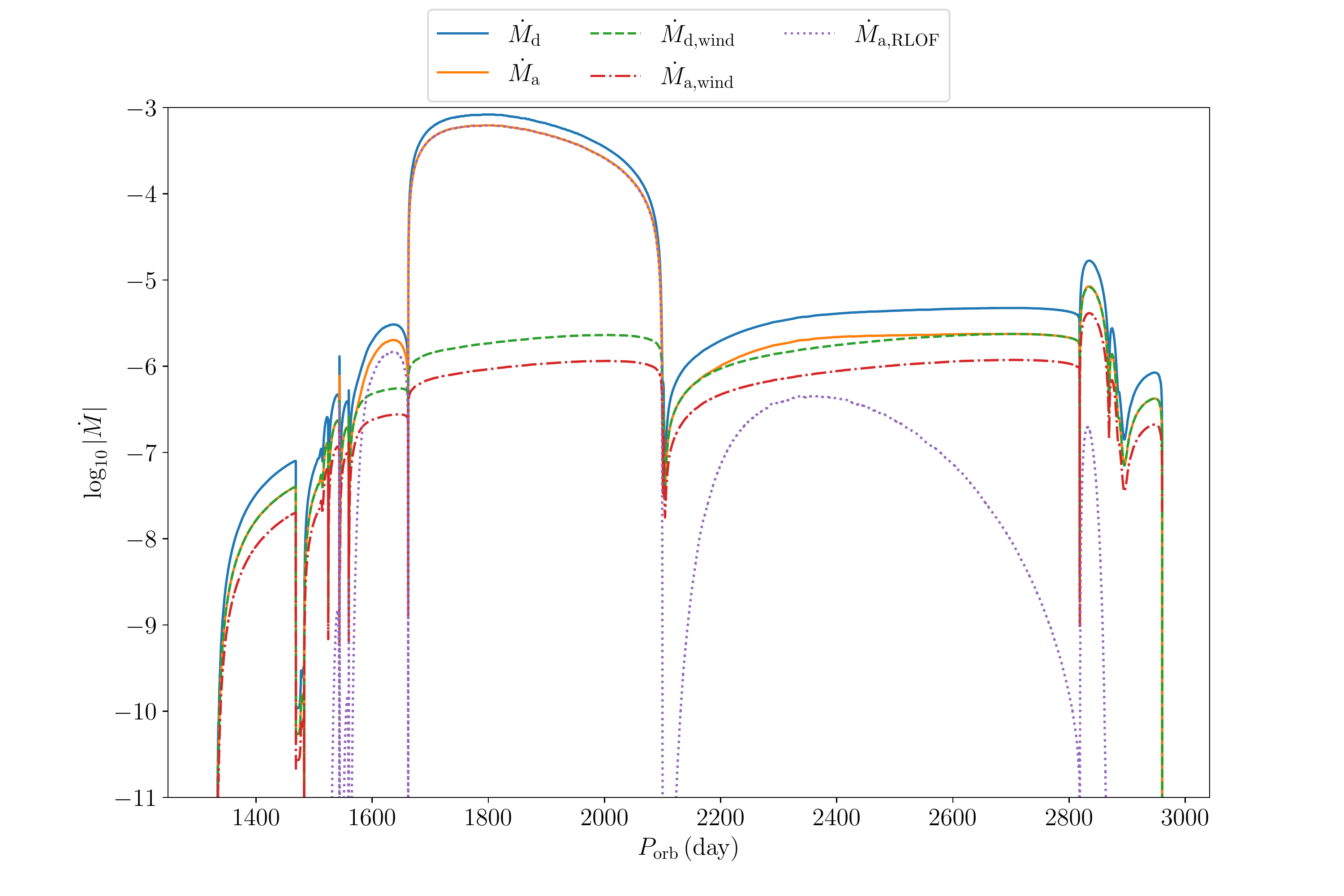}
\caption{The plot shows total $\dot{M}$ with respect to $P_{\rm orb}$ for the donor star (blue) and the accretor star (orange) for the BF model (with the inclusion of the WRLOF). The green dashed line represents the wind mass loss from the donor star. The red dash-dotted line is the wind accretion rate onto the accretor. The purple dashed line is the MT rate only by regular RLOF. The initial $P_{\rm orb}$ is 1330 days.}
\label{Mdot_P_wRLOF_v3.pdf}
\end{figure*}

Figure \ref{Mdot_P_wRLOF_v3.pdf} provides detailed information on the MT within the BF model. This figure gives the total mass loss rate from the donor star $\dot{M}_{\rm d}$ in blue and the total mass accretion rate of the accretor star $\dot{M}_{\rm a}$ in orange, both as a function of orbital period during the binary evolution. The WRLOF mass loss rate from the donor $\dot{M}_{\rm d,\,wind}$ is shown as the green dashed line. The WRLOF accretion rate onto the accretor $\dot{M}_{\rm a,\,wind}$ is shown as the red dash-dot line. The regular RLOF mass accretion rate onto the accretor $\dot{M}_{\rm a,\,RLOF}$ is the purple dotted line.  The relations between these rates are as follows

\begin{align*}
\dot{M}_{\rm d} = \dot{M}_{\rm d,\,wind} + \dot{M}_{\rm d,\,RLOF}, \\
\dot{M}_{\rm a} = \dot{M}_{\rm a,\,wind} + \dot{M}_{\rm a,\,RLOF}, \\
\dot{M}_{\rm a,\,wind} = \beta_{\rm acc} \dot{M}_{\rm d,\,wind}, \\
\dot{M}_{\rm a,\,RLOF} = \beta \dot{M}_{\rm d,\,RLOF},
\end{align*}
where $\beta$ is the RLOF MT efficiency. 

\begin{figure}[tp]
\includegraphics[width=0.5\textwidth]{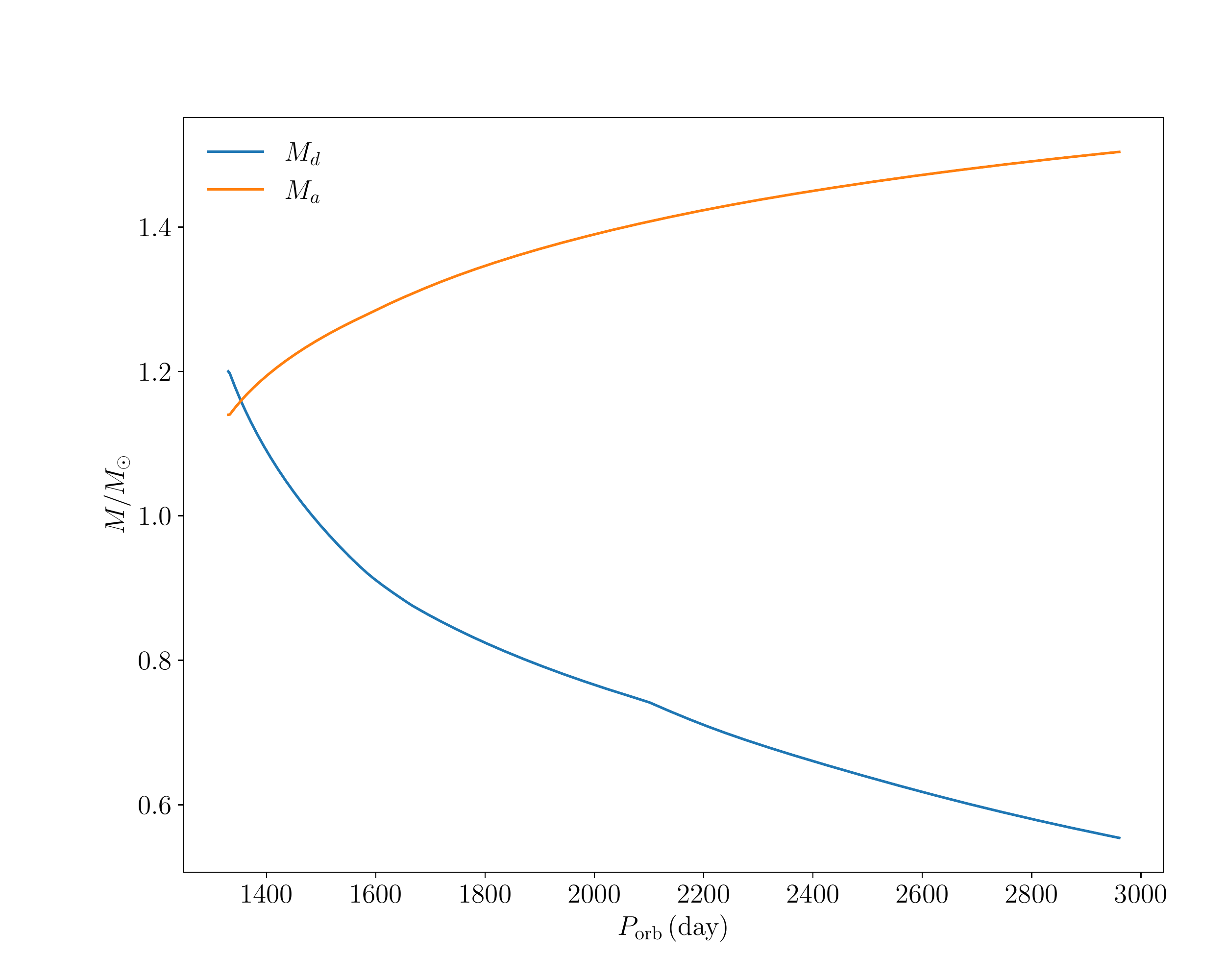}
\caption{This plot gives the mass of the donor star (blue) and accretor star (orange) with evolution of the system.}
\label{Mass_P_wRLOF.pdf}
\end{figure}

In Figure \ref{Mdot_P_wRLOF_v3.pdf}, the accretor star begins to accept material at an initial period of 1330 days, when the donor star is in the RGB phase of evolution. The first bump ends at an orbital period of 1470 days, when the donor star finishes its RGB evolution (6.2 Gyr). The donor star is entirely within its Roche-lobe radius, so this first MT results only from the stellar wind of the donor star, at a MT transfer efficiency of $50\%$ (see Section \ref{sec:wind MT}).
Figure \ref{Mass_P_wRLOF.pdf} shows the consequent mass changes of the donor (blue solid line) and accretor star (orange solid line) as a function of the system orbital period. During the donor's RGB mass loss, the donor mass $M_{\rm d}$ drops from $1.20\,M_{\odot}$ to $1.01\,M_{\odot}$ and the accretor mass $M_{\rm a}$ increases from $1.14\,M_{\odot}$ to $1.24\,M_{\odot}$. From here on in the evolution the accretor is the more massive star.

\begin{figure}[tp]
\includegraphics[width=0.5\textwidth]{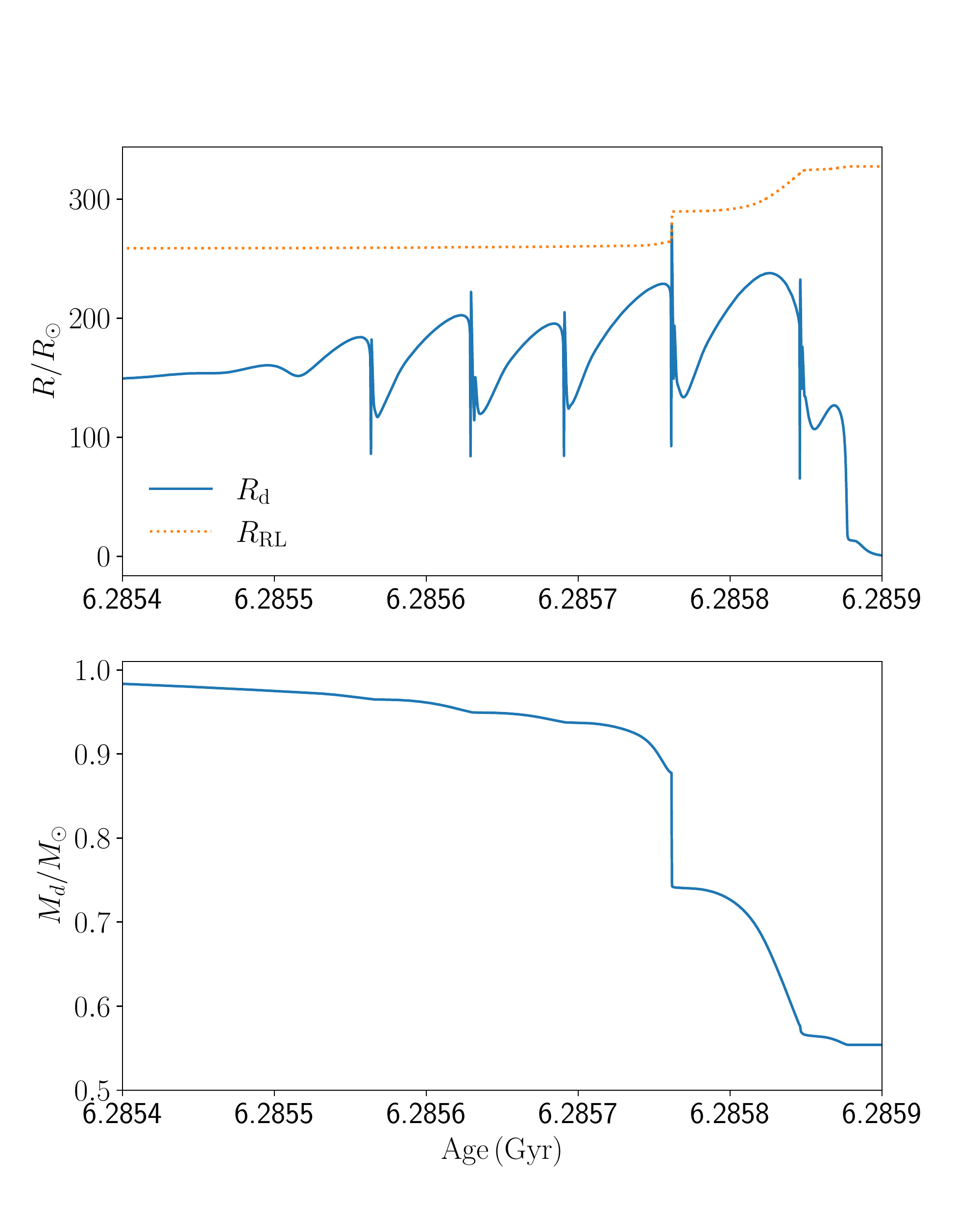}
\caption{The upper(lower) plot gives $R_{\rm d}$($M_{\rm d}$) as a function of time during the thermal pulses in the AGB phase. The upper panel also gives the $R_{\rm RL}$ as a function of age in the dashed orange line.}
\label{R1_vs_t.pdf}
\end{figure}

The next series of bumps and spikes in the $\dot{M}$ versus $P_{\rm orb}$ plane begins after $P_{\rm orb}=1483$ days and is enhanced by AGB thermal pulses. We show the donor radius, the Roche-lobe radius and the mass changes of the donor during these thermal pulses in Figure \ref{R1_vs_t.pdf}. During this short timescale (0.4 Myr), the donor star loses a significant amount of mass through a combination of regular RLOF and WRLOF, decreasing from $0.97\,M_{\odot}$ to $0.55\,M_{\odot}$. The largest spike in donor radius, increasing to 278 $R_{\odot}$ (top panel of Figure \ref{R1_vs_t.pdf}), corresponds to the largest MT rate ($\dot{M}\sim 10^{-3}\dot{M}_{\odot}/{\rm yr}$) in Figure \ref{Mdot_P_wRLOF_v3.pdf} (starting at $P_{\rm orb}=1663$ days). 

Only during this largest thermal pulse does the donor radius exceed the Roche-lobe radius. Therefore, only during the largest thermal pulse is the total MT rate dominated by regular RLOF. The mass drops from 0.88 to 0.74 $M_{\odot}$. At all other times the WRLOF is the primary contributor to the total MT rate.

After losing most of the envelope, the donor star cannot maintain the helium burning shell and shrinks in radius. The MT turns off.

The total mass transferred from the donor to the accretor is 0.36 $M_{\odot}$. In Table 1, we show the mass gains of the BSS from each of the three processes: RGB WRLOF, AGB WRLOF, and regular RLOF. Their contributions are comparable, but the largest contribution is from AGB WRLOF.

\subsection{Evolution of Stellar Interiors during MT in the Best-Fit Model}
\label{sec: bi_evol}

\begin{figure}[tp]
\includegraphics[width=0.5\textwidth]{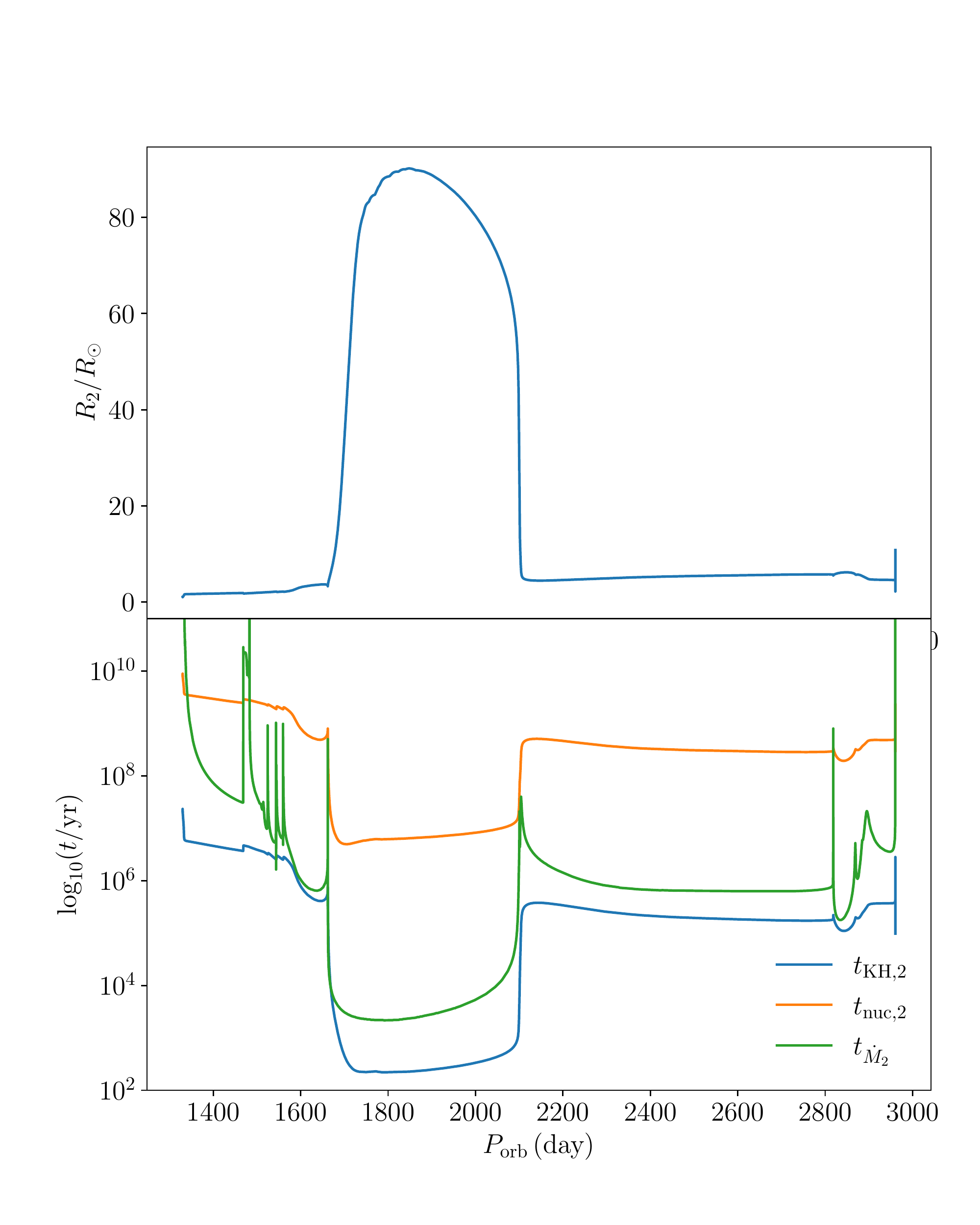}
\caption{Top: The radius of the accretor star as a function of $\log_{10} P_{\rm orb}$. Bottom: Three timescales - thermal $t_{\rm KH,2}$, nuclear $t_{\rm nuc,2}$, and accretion $t_{\dot{M}_2}$ - of the accretor in terms of $\log_{10} P_{\rm orb}$.}
\label{accr_kipp.pdf}
\end{figure}

Figure \ref{accr_kipp.pdf} shows the evolution of the photospheric radius and of three physical timescales for the accretor as a function of $P_{\rm orb}$, in the middle and bottom panels respectively. 

The radius increase starting at $P_{\rm orb}=1663$ days corresponds to the highest bump in Figure \ref{Mdot_P_wRLOF_v3.pdf}, where the MT rate reaches $10^{-3}\,M_{\odot}/{\rm yr}$.
At the beginning of this event, the thermal and MT timescales are comparable. With the rapid increase in MT at the beginning of the pulse, the MT timescale becomes very short.
The accretor is unable to thermally restructure itself, and responds with a rapid and very large increase in radius. With the associated decrease in the thermal timescale, the high MT and large radius continue throughout the pulse. With the end of the pulse and the decrease in MT rate, the star returns to approximately its prior radius. (The later phase of accretor radius growth is after the MT has stopped, when the BSS star enters the RGB phase.)

\begin{figure}[tp]
\includegraphics[width=0.5\textwidth]{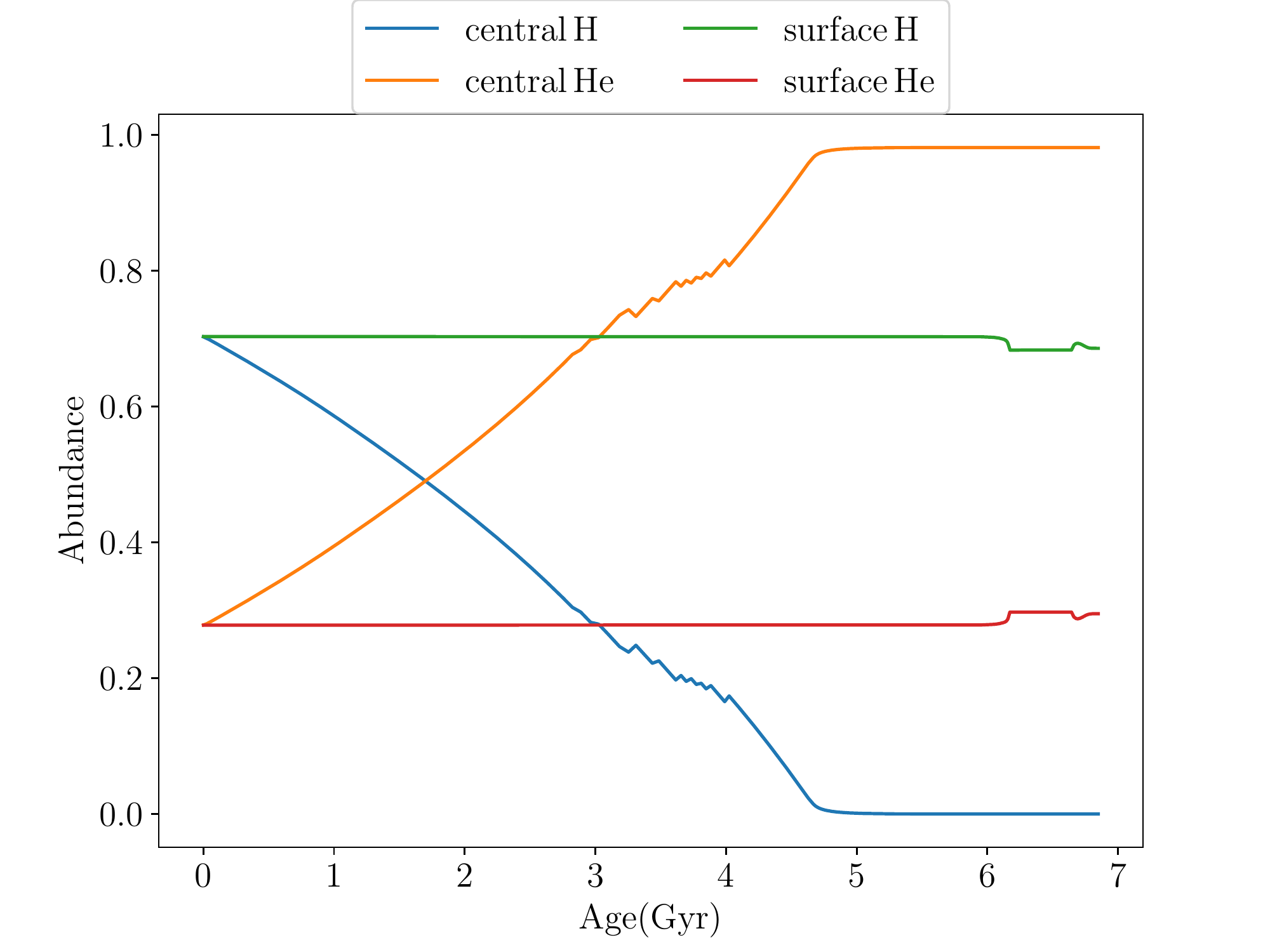}
\caption{The abundance profile as a function of system age for the accretor star.}
\label{accr_abund_pf.pdf}
\end{figure}

Figure \ref{accr_abund_pf.pdf} shows the interior elemental abundances of the accretor star as a function of age. Before the MT begins, the hydrogen and helium abundance evolutions reflect the standard core burning of a 1.2 $M_{\odot}$ star. When the WRLOF MT transfer from the RGB donor begins at 5.6 Gyr, the core burning of the accretor star has ended, so the central abundance of the accretor does not change further. However, at 6.2 Gyr the surface hydrogen drops and surface helium increases as the accretor star accepts helium enriched material from the AGB donor star. 

\begin{figure}[tp]
\includegraphics[width=0.5\textwidth]{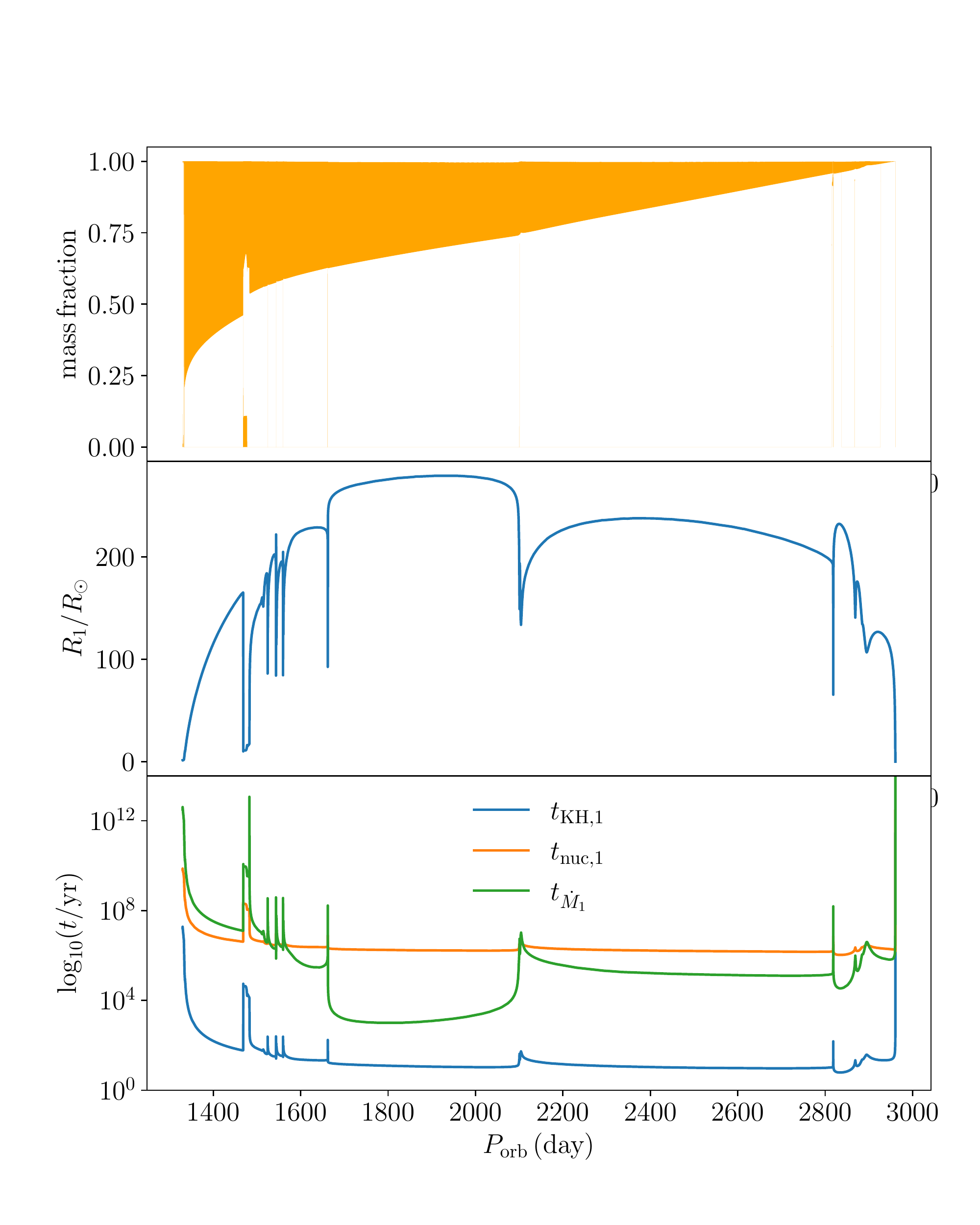}
\caption{Top: The mass fraction and the location of the convective zone of the donor star versus $P_{\rm orb}$. Middle: The radius of the donor star as a function of $P_{\rm orb}$. Bottom: Three timescales - thermal $t_{\rm KH,1}$, nuclear $t_{\rm nuc,1}$, and accretion $t_{\dot{M}_1}$ - of the donor star in terms of $P_{\rm orb}$.}
\label{donor_kipp.pdf}
\end{figure}

Figure \ref{donor_kipp.pdf} shows the donor star convective zones (as mass fractions), the photosphere radius and three timescales, all as a function of $P_{\rm orb}$, in the top, middle and bottom panels, respectively. At the ZAMS phase, the donor star has a tiny convective core and a deep convective envelope. The mass of this convective envelope decreases throughout due to wind mass loss and, during a thermal pulse, by regular RLOF. Near $P_{\rm orb}=1470$ days, the donor star finishes RGB evolution and core helium is ignited, initiating a convective core and causing the star to shrink in radius. After the completion of the core helium burning, the core convective zone disappears, shell helium burning starts and the star enters the AGB phase, where the star still has a radiative core and a convective envelope. On the second panel, the largest radius of the donor star during the biggest thermal pulses can reach $\sim 275\,R_{\odot}$. In the bottom panel, during the AGB phase the MT timescale is shorter than the nuclear timescale, which means the regular RLOF is thermally driven.

\begin{figure}[tp]
\includegraphics[width=0.5\textwidth]{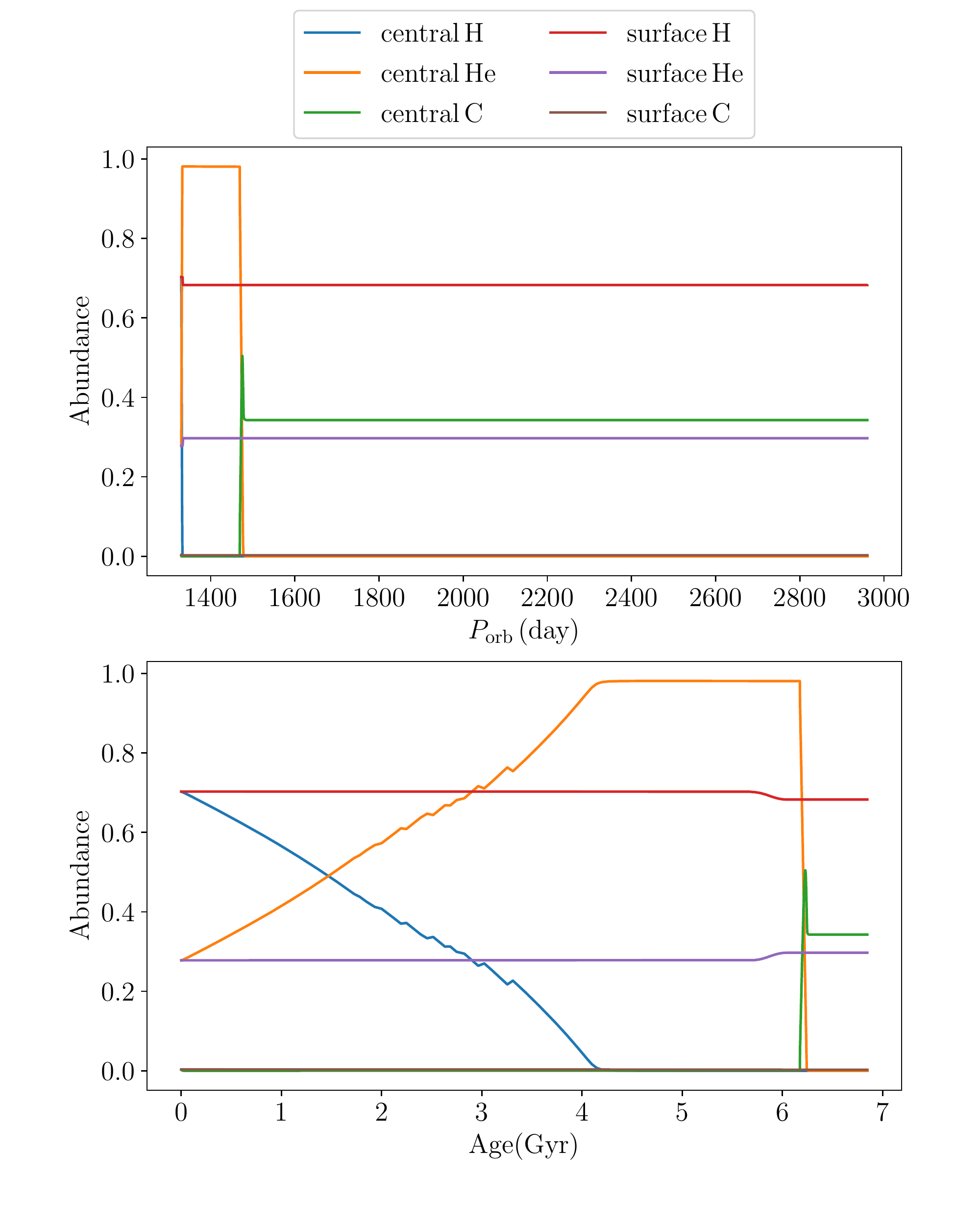}
\caption{{The abundance profile as a function of $P_{\rm orb}$ and system age for the donor star.} }
\label{donor_abund_pf.pdf}
\end{figure}

Figure \ref{donor_abund_pf.pdf} gives the central and surface hydrogen, helium and carbon mass fractions of the donor star as a function of $P_{\rm orb}$ (top panel) and age (bottom panel). The donor star is initially more massive, so core hydrogen exhaustion occurs at 4.3 Gyr, earlier than the accretor star. The central helium ignition is near 6 Gyr and marks the end of the RGB phase for the donor star. The central carbon rises to 0.5 in mass fraction, and then drops to 0.34 as carbon converts to oxygen in the center. At this time, central helium is exhausted as well.

\section{RLOF-Only Models}
\label{sec: RLOF-Only Model}

The long orbital period, the large BSS luminosity (and presumably mass gain), and the CO WD indicating an AGB MT origin all suggest that WOCS 4540 may be a likely case involving WRLOF. Indeed, our simulations producing the BF model for WOCS 4540 have found WRLOF to be the primary contributor to the BSS mass gain (Table 1). WRLOF is a relatively new addition to the portfolio of MT processes. In addition, its efficacy has had limited theoretical development or constraint from observations. As such, Occam's Razor suggests that we also explore whether regular RLOF alone can produce WOCS 4540.


\begin{figure}[tp]
\includegraphics[width=0.5\textwidth]{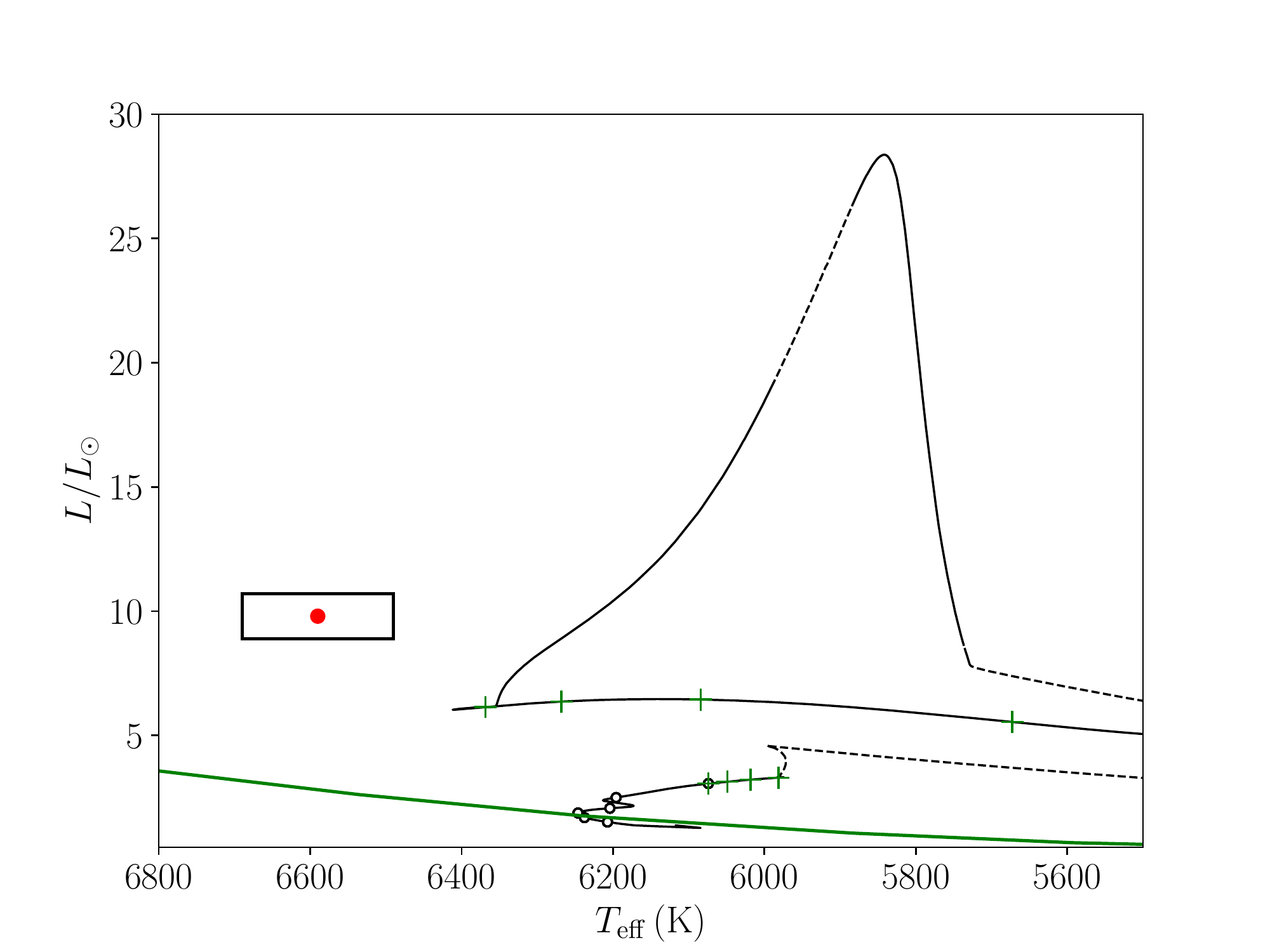}
\caption{The HR diagram for the evolution of the secondary star into a BSS for a selected model including only regular RLOF and the standard wind scaling factors. Otherwise the figure is the same as Figure \ref{HR2_zoomin.pdf}.}
\label{HR2_zoomin_failed.pdf}
\end{figure}

We first consider a model that 
starts with the same initial parameters as the BF model. The only differences are that (a) WRLOF is not applied during the simulation and (b) the regular RLOF is fully conservative (MT efficiency of 100\%). Figure \ref{HR2_zoomin_failed.pdf} shows the evolutionary track of the accretor star from the ZAMS to the post-MT phase. Before MT, the evolution of the star is the same as in the BF model. After MT, the accretor star never reaches the observed values of WOCS 4540 in either $T_{\rm eff}$ and luminosity.

\begin{figure}[tp]
\includegraphics[width=0.5\textwidth]{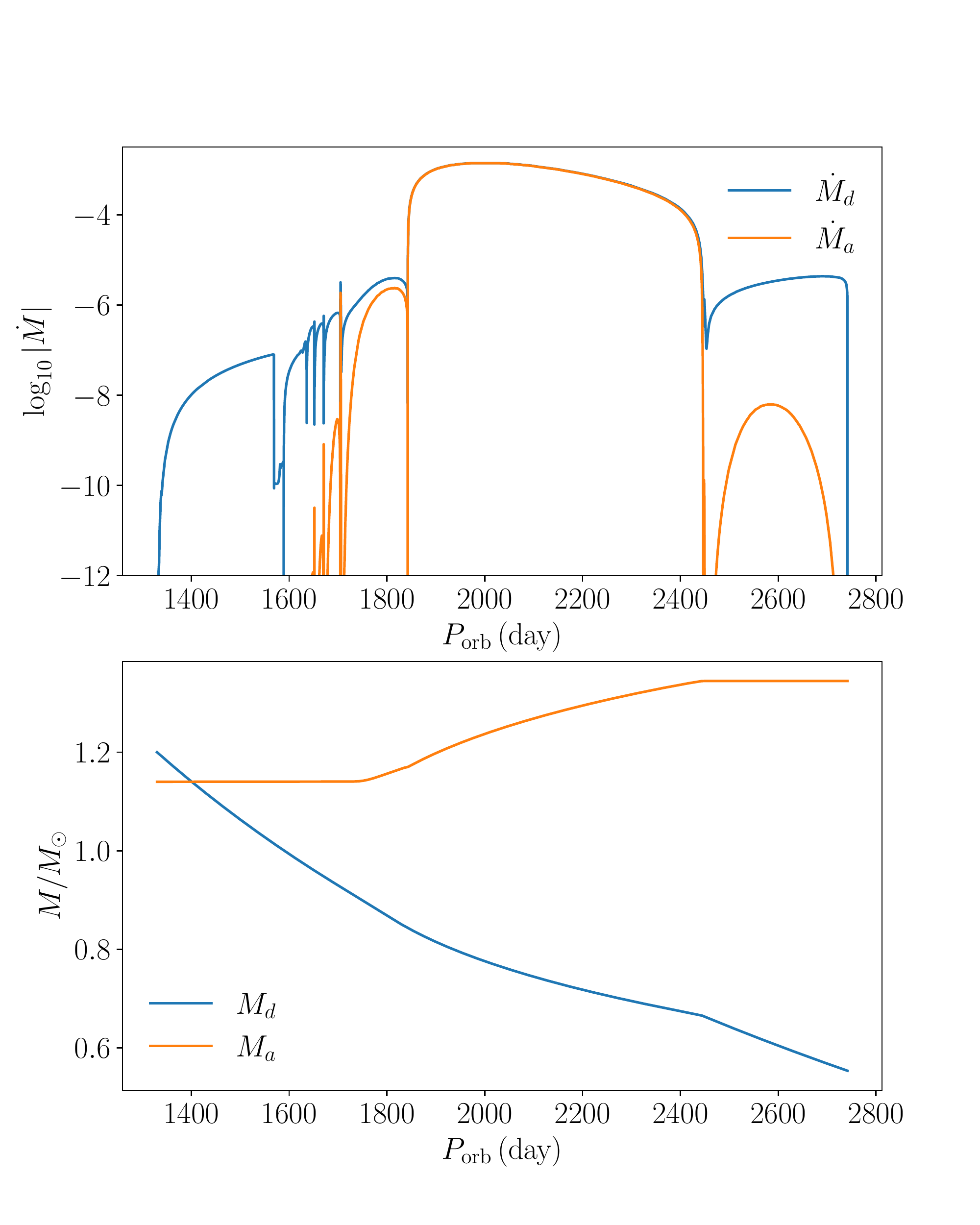}
\caption{The top(bottom) panel follows the description of Figure \ref{Mdot_P_wRLOF_v3.pdf}(\ref{Mass_P_wRLOF.pdf}), but for a model including only regular RLOF and the standard wind scaling factors.}
\label{Mdot_P_failed.pdf}
\end{figure}

Figure \ref{Mdot_P_failed.pdf} shows the MT history of this model. The top panel shows the mass loss rate from the donor star in blue and the mass accretion rate onto the accretor star in orange, as a function of $P_{\rm orb}$. There is no MT when the donor is in the RGB phase, the horizontal branch phase and the beginning of the AGB phase ($P_{\rm orb}$ between 1330 to 1635 days). MT starts when the donor star begins its thermal pulses. 

The bottom panel shows the evolution of the masses of the two stars (donor mass in blue and accretor mass in orange). In the RGB phase, the donor star drops from 1.2 $M_{\odot}$ to 1.01 $M_{\odot}$ through wind mass loss, as in the BF model. From the beginning of the AGB phase to the first thermal pulse, the donor mass drops further from 1.01 $M_{\odot}$ to 0.96 $M_{\odot}$. During the MT phase, the donor star mass drops from 0.96 $M_{\odot}$ to 0.55 $M_{\odot}$ and then becomes a WD. The accretor star gains mass from 1.14 $M_{\odot}$ to 1.34 $M_{\odot}$. This lower final mass is the essential reason that the acccretor star cannot match the BSS effective temperature or luminosity.

The fundamental challenge to this model is the wind mass loss of the donor, which depletes the donor's envelope mass reservoir. In the BF model a substantial fraction of this wind mass loss is accreted by the companion through WRLOF. Without WRLOF, the donor wind mass is lost from the system.

We thus explore whether reducing the donor wind mass loss can produce WOCS 4540. While challenging MT efficiency and numerical stability, a model with MT only through regular RLOF was able to do so. We call this model the RLOF-Only model.

For the accretor to reach the observed luminosity of the BSS, it was necessary to significantly lower the stellar wind efficiency factors, with the Reimer scaling factor at 0.1 and the Bl\"{o}cker scaling factor at 0.07 (compared to the MESA default values of 0.5 and 0.1, respectively, used in the BF model). The reduced Reimer factor is at the lower limit of the range found by \citet{2012MNRAS.419.2077M} via astroseismology analysis of RGB stars in the open clusters NGC 6791 and NGC 6819. The reduced Bl\"{o}cker scaling factor is in between the MESA default and the calibration by \citet{2000A&A...363..605V} from fitting the lithium luminosity function in the Large Magellanic Cloud.

\begin{figure}[tp]
\includegraphics[width=0.5\textwidth]{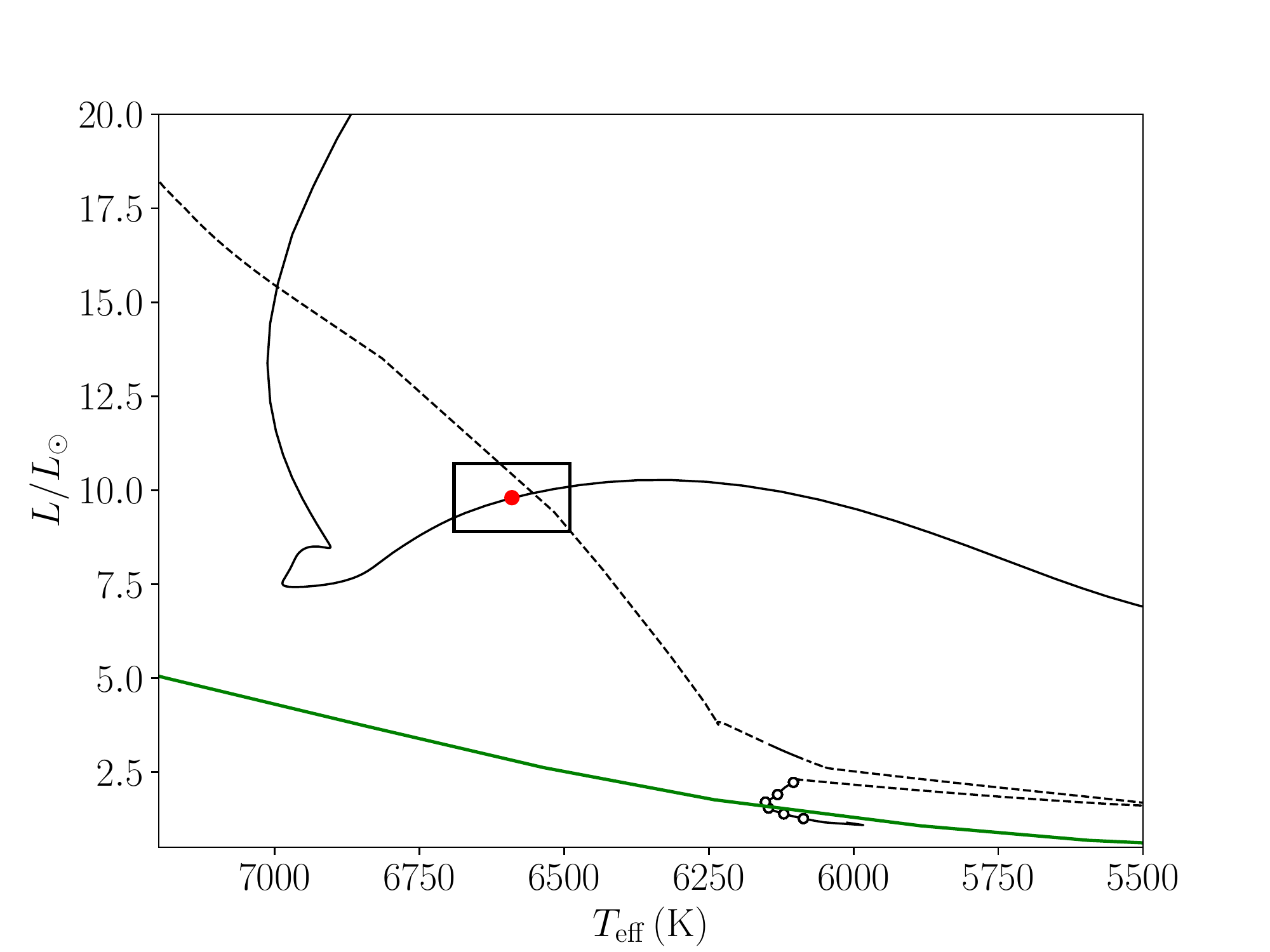}
\caption{The HR diagram for the evolution of the secondary star into a BSS for the RLOF-Only Model. Otherwise the figure is the same as Figure \ref{HR2_zoomin.pdf}.}
\label{HR2_zoomin_RLOF.pdf}
\end{figure}

The RLOF-Only model starts with a 1.2 $M_{\odot}$ donor star, an accretor mass of 1.1 $M_{\odot}$ and an orbital period of 1500 days. Figure \ref{HR2_zoomin_RLOF.pdf} shows the HR diagram for the accretor star. The evolution begins on the ZAMS at $(T_{\rm eff}/{\rm K},\, L/L_{\odot})=(5985, 1.01)$. The mass accretion starts after 6 Gyr when the accretor star is exhausted in central hydrogen, shown as the dashed line. The AGB thermal pulses of the donor star initiate the regular RLOF, resulting again in a large luminosity increase during the MT phase. For this model, the maximum luminosity reaches 350 $L_{\odot}$, which is not shown in Figure \ref{HR2_zoomin_RLOF.pdf}. After the donor star exhausts its shell helium, MT stops. At this point the accretor star has become a BSS, and its subsequent evolution passes through the observed position on the CMD. The donor star has become a CO WD near the observed position on the CMD as well. 

\begin{figure}[tp]
\includegraphics[width=0.5\textwidth]{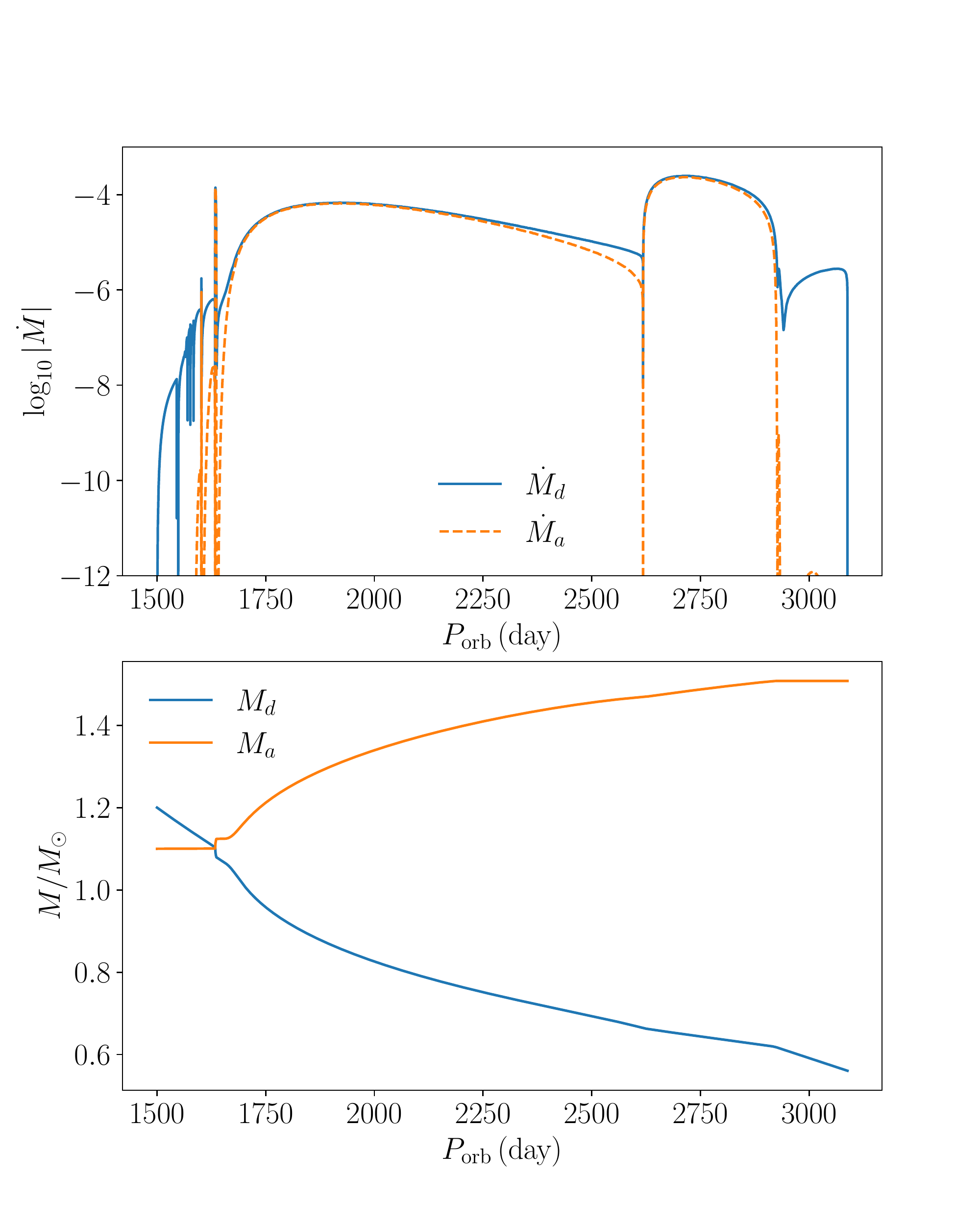}
\caption{The figure follows the description of Figure \ref{Mdot_P_failed.pdf}, but for the RLOF-Only Model.}
\label{Mdot_P.pdf}
\end{figure}

Similar to Figure \ref{Mdot_P_wRLOF_v3.pdf}, Figure \ref{Mdot_P.pdf} shows the detailed MT for the RLOF-Only Model. At $P_{\rm orb}=1546$ days, the donor finishes the RGB phase evolution with a mass of $1.16\,M_{\odot}$, somewhat reduced due to stellar wind mass loss. The donor star enters the AGB phase at $P_{\rm orb}=1549$ days. The regular RLOF begins after the donor star has several thermal pulses, at $P_{\rm orb}=1590$ days.

\begin{figure}[tp]
\includegraphics[width=0.5\textwidth]{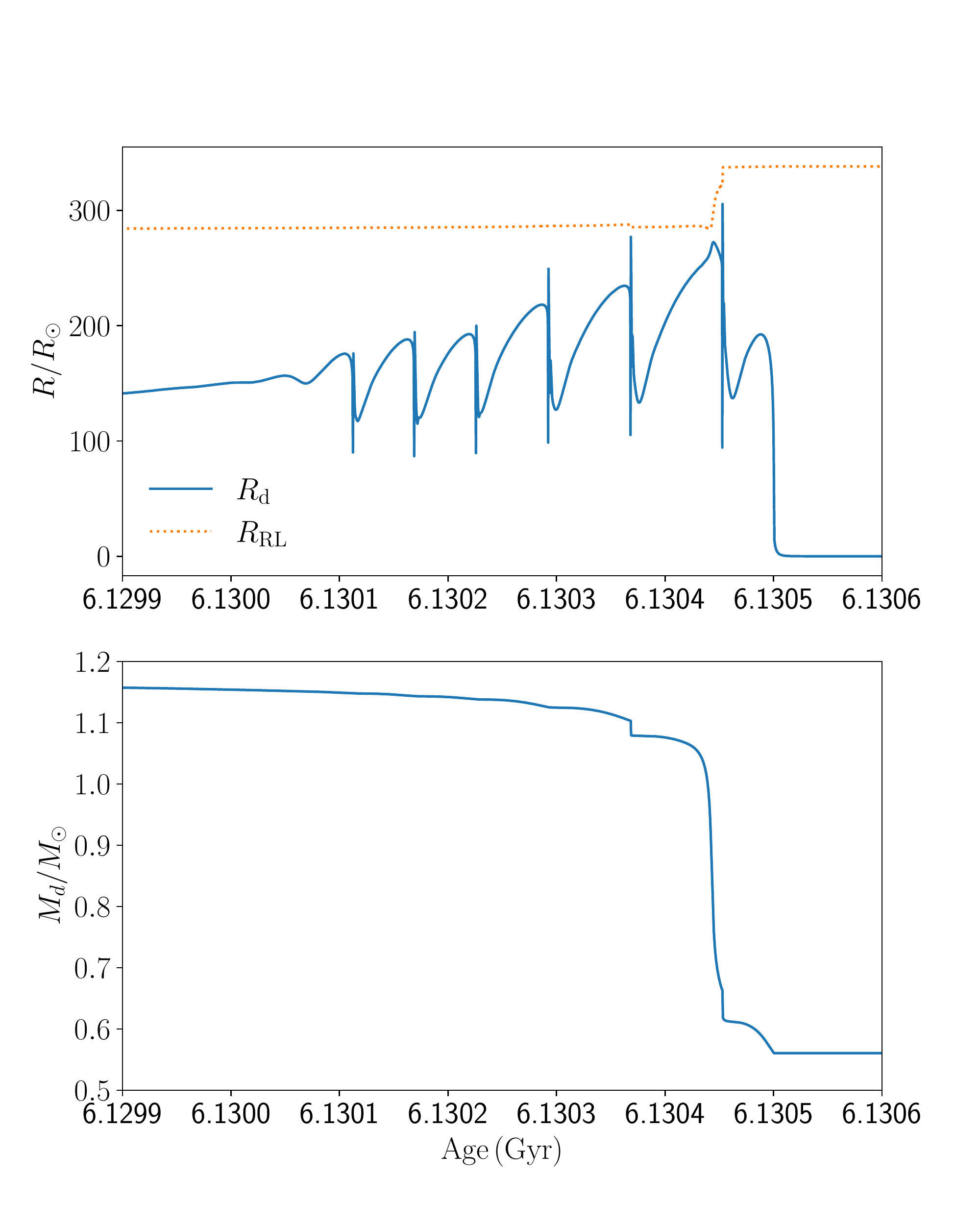}
\caption{Similar to the figure description of Figure \ref{R1_vs_t.pdf}, but for the RLOF-Only Model.}
\label{R1_vs_t_nowind.pdf}
\end{figure}

Figure \ref{R1_vs_t_nowind.pdf} provides detailed insight into the stellar and Roche lobe radii during the donor's AGB thermal pulses phase. The top panel shows that regular RLOF occurs at the second (6.1304 Gyr) and the third last (6.1305 Gyr) thermal pulses, where the radius of the donor star is very close to its Roche lobe radius. These phases correspond to the largest MT rates in Figure \ref{Mdot_P.pdf}. At the second to last thermal pulse, the star radius reaches 300 $R_{\odot}$. The second panel of Figure \ref{R1_vs_t_nowind.pdf} shows the mass change of the donor star. Before thermal pulses, the donor star mass is 1.15 $M_{\odot}$. At the third to last thermal pulse, donor's mass decreases from 1.11 to 1.08 $M_{\odot}$. At the second to last thermal pulse, the donor star mass drops from 1.0 to 0.61 $M_{\odot}$.

To increase the accretor luminosity (and associated mass) to match WOCS 4540, 
the regular RLOF MT efficiency must be 100\%; all the material lost by the donor star through RLOF must be accepted by the accretor. Equally important, the reduction in the wind mass loss efficiencies is required in order to retain more mass in the envelope as a reservoir for the regular RLOF. 

{Because of the donor's additional mass loss through stellar wind, which increases the specific angular momentum of the system, the binary separation keeps increasing. As a consequence, the regular RLOF MT is stable throughout. Additionally, the increase in mass transferred after mass reversal increases the final orbital period to nearer the observed 3030 days.}

To summarize the key finding of this comparison, for the RLOF-Only Model to produce a $\sim 1.5\,M_{\odot}$ BSS,  fully conservative regular RLOF MT is required. Thus, a $\sim 1.5\,M_{\odot}$ BSS is at the mass upper limit of BSS formation in NGC 188 with a RLOF-Only Model. (We note that increasing the initial accretor mass leads to numerical instabilities and unsuccessful models.) This is to be compared to the BF model with a regular RLOF MT efficiency of 75\%, which is supplemented by WRLOF.

\section{Discussion}
\label{sec:Discussion}

\subsection{Accretor Luminosity Burst During the AGB MT}
The luminosity of the accretor star varies significantly throughout the MT, and most dramatically during the donor's biggest thermal pulse where the radius of the donor star reaches $\sim 275\,R_{\odot}$ and the MT rate is at its peak. The corresponding luminosity of the accretor star reaches $2500 L_{\odot}$, although the duration of the luminosity boost is as short as thousands of years. 

\begin{figure}[tp]
\includegraphics[width=0.5\textwidth]{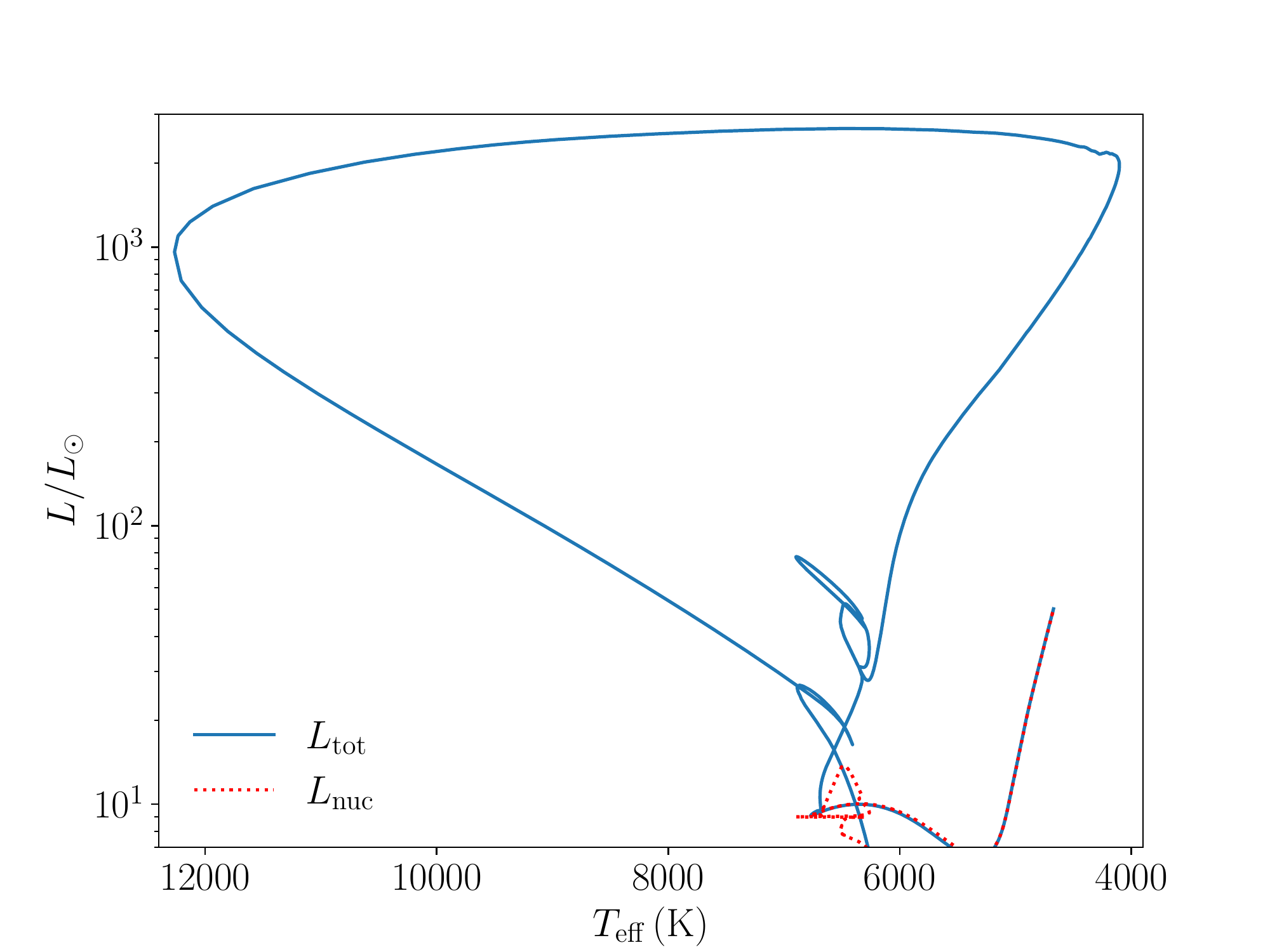}
\caption{The bolometric luminosity of the accretor star (blue solid line) and its  nuclear luminosity (red dotted line).}
\label{checkL_teff4540accr.pdf}
\end{figure}

Figure \ref{checkL_teff4540accr.pdf} compares the bolometric photosphere luminosity calculated by MESA (blue solid line) and the luminosity supported by nuclear reactions (red dotted line) for the accretor's regular RLOF accretion phase and post-MT phase. Before and after the MT phase, the entire luminosity of the star is produced by nuclear reactions. From $T_{\rm eff,2}=4600\sim 6700$ K (post-MT evolution), the red dotted line and the blue solid line overlap.  


MESA calculates the stellar structure during accretion by introducing an extra term for the gravitational heating rate in the energy equation. The gravitational heating rate is a function of the mass accretion rate. (See Section 5.1 of the MESA second instrument paper by \citealt{2013ApJS..208....4P} and Section 7 of the MESA third instrument paper by \citealt{2015ApJS..220...15P} and the discussions.) Note that this treatment does not add the accretion luminosity ($L_{\rm acc}$ estimated by $(GM_2/R_2)\dot{M}_2$) into the total bolometric luminosity. 



Comparing the evolution of the BSSs in WOCS 5379 and WOCS 4540, in WOCS 5379 the BSS never reaches a high $T_{\rm eff}$ ($T_{\rm eff,max} \sim 6400$ K) or large radius. In the stable RLOF phase, the BSS in WOCS 5379 gains mass from 1.1 $M_{\odot}$ to 1.2 $M_{\odot}$. The radius increases initially. After it reaches 1.6 $R_{\odot}$, a central convective core appears and the radius drops. On the other hand the $T_{\rm eff}$ and radius of the WOCS 4540 BSS reach very large values in the rapid RLOF phase during the largest thermal pulse, yielding a very large bolometric luminosity. In at least some cases, stars cannot respond to the rapid accretion during the largest thermal pulse.

\subsection{Stellar Rotation}

The BSS in WOCS 4540 has a rotation period of $1.8^{+2.3}_{-0.92}$ days. This rotation is very fast compared with MS stars in NGC 188. In addition to WOCS 4540, \citet{2018ApJ...869L..29L} show the rotation periods of 12 BSS and of MS stars in binary systems containing a WD, both in NGC 188 and the field, as a function of WD cooling age. The MS stars with WD ages less than 100 Myr are very rapidly rotating (e.g, rotation periods less than 1 day), while the somewhat older BSSs decrease in spin rate up with increasing age up to an age of 1 Gyr. The youngest of these systems, including WOCS 4540, are valuable examples for studying how MT spins up the accretor star. 

\citet{1981A&A...102...17P} and \citet{2017A&A...606A.137M} argue that the accretor star can be accelerated to the critical rotation velocity with accretion of only a few percent of its total mass.  \citet{2017A&A...606A.137M} find this to be problematic for CEMP-{\it s} dwarts, since the progenitor stars cannot accrete enough mass to explain their chemical properties before excessive spin-up.

\citet{1975ApJ...198..383L} estimate the specific angular momentum that can be transferred onto an accretor through MT via an accretion disk or direct impact of the accretion stream.
Implementing the formalism of \citet{1975ApJ...198..383L} in fast binary-population simulations and in detailed binary evolution codes, spin angular momentum evolution through MT has been explored in massive binaries \citep{2013ApJ...764..166D,2021ApJ...923..277R}, where the models and observations agree well in general.

In this work, the evolution of stellar rotation has not been incorporated in the simulation. Even so, the BF model indicates that throughout its evolution, WOCS 4540 is always in a wide orbital system (orbital periods from 1330 days to 2961 days). Although the the accretor star of the WOCS 4540 system is a fast rotator, its spin angular momentum is still negligible compare with the system orbital angular momentum. Therefore, ignoring the spin of the two stars in this work would not change the resulting orbital evolution. 

{On the other hand, considering stellar evolution, fast rotation could lead to efficient rotational mixing, which affects the main-sequence lifetime of the stars (burning more hydrogen fuel into helium). 
Another effect caused by rotation is meridional circulation, where the chemical abundance could change even if the rotation is not fast. Incorporating stellar rotation is much needed in future models of these systems.}

\subsection{Alternative Formation Pathways}

Motivated by the very young CO WD, this paper has explored in detail the formation and evolution of the WOCS 4540 system via MT. While this process has proven successful, there is not much latitude in the process to create yet more luminous BSSs, a few of which in fact are present in NGC 188. Others, e.g. \citealt{2021ApJ...908..229L}, have suggested that the most luminous BSSs may be the result of efficient mergers.

In this context, the evolutionary role of the current WD is perhaps rather different, being the tertiary of an initially triple system. In the context of triple systems, the role of the tertiary has been argued to drive a merger via Kozai-Lidov processes (e.g., \citealt{2009ApJ...697.1048P}). However, in such a scenario the tertiary being a very young CO WD would be merely serendipitous.

Alternatively, the tertiary may have just recently evolved. \citealt{2019ApJ...876L..33P} have explored the evolution of such a triple system, finding that in certain configurations mass lost from the tertiary form a circumbinary disk around the inner binary. In their particular simulation they find MT to occur onto both inner stars, forming two BSSs in a short-period system (as is in fact observed in WOCS 7782 of NGC 188). However, circumbinary disks can also act to shrink inner binaries (e.g., \citealt{1994ApJ...421..651A,2020MNRAS.496.1819L}), possibly into a merger.

Finally, we note that within a cluster environment there is always the possibility that a current binary did not form with the cluster as such, but rather is the result of dynamical exchanges. Dynamical exchanges favor forming systems with the most massive objects in the cluster, such as the BSSs. It is relatively unlikely that such a low-mass WD be exchanged into a binary containing the BSS and a very low-mass companion (now ejected; N. Leigh, private communication). Arguably, the WOCS 4540 BSS could have been exchanged into another binary including the WD and a yet lower mass companion. A WD of such low-mass can perhaps be formed by a turnoff star in NGC 188 \citep{2008MNRAS.387.1693C}, although it would be rather fortuitous if the exchange happened only 100 Myr thereafter. Finally, we note that WOCS 4540 lies in projection at approximately one core radius, also not suggestive of a recent resonant exchange encounter.

\section{Conclusion and Future Research}
\label{conclusion}

WOCS 4540 is the longest orbital period blue straggler - white dwarf pair in the old open cluster NGC 188. It also contains one of the most luminous BSS in the cluster. Combining the orbital data with extensive data for both of the stars and the cluster, including especially the mass determination of the WD companion, the system has ample information for detailed modeling of its formation and evolution. We use the MESA binary module to simulate the system, with our incorporation of the wind Roche-lobe overflow formalism for wind accretion.

Our model starts with two ZAMS stars, with masses of 1.2 $M_{\odot}$ and 1.14 $M_{\odot}$ in an orbital period of 1330 days. The massive star evolves first, and begins MT via the wind during the RGB phase. This wind MT continues as WRLOF throughout the AGB phase. The donor fills its Roche-lobe radius only during its AGB phase, where regular RLOF occurs only during the largest thermal pulse of the donor. The donor's radius grows beyond 200 $R_{\odot}$ and the maximum mass loss rate of the donor star then reaches $10^{-3}\,M_{\odot}/{\rm yr}$. 

Upon completion, the simulation donor star becomes a 0.55 $M_{\odot}$ C-O core WD, and the accretor star becomes a 1.5 $M_{\odot}$ blue straggler. The model system has a very long orbital period of 2961 days.

The BF model matches most of the stellar and orbital parameters derived from observations of WOCS 4540 (see Table \ref{observation vs model}). The simulation adds to the observation knowledge base the current mass of the BSS and its transformation age (\citealt{2021ApJ...908....7S}) of only 120 Myr. This BSS formed very recently. Notably not reproduced by these simulations are the orbital eccentricity and BSS spin, neither of which are included in the simulation. 

In addition to the success of the MT simulation, the astrophysical findings of this investigation of WOCS 4540 are the following:

\bigskip

1) In the BF model, three MT processes contribute roughly equally to the 0.36 $M_{\odot}$ gained by the accretor: RGB WRLOF, AGB WRLOF and regular RLOF during the largest AGB thermal pulse. The overall MT efficiency is 55\%; the required MT efficiency for the regular RLOF is 75\%.

\bigskip

2) With standard wind mass loss parameters, WOCS 4540 cannot be produced by regular RLOF alone; the wind mass loss from the donor so depletes the donor envelope that insufficient mass remains to yield a BSS of adequate luminosity.

To create the BSS of WOCS 4540 with only regular RLOF requires Reimer and Bl\"{o}cker factors to describe the strength of the wind that are at the lower side of observed calibrations. More challenging, even with reduced wind depletion of the envelope, fully conservative MT is needed to explain such a luminous BSS. In addition to being physically implausible, such a high MT efficiency challenged numerical (and perhaps physical) stability. Even with these challenges, a 1.5 $M_{\odot}$ BS at 3000 day is at the very upper limit of the produceable BSS mass. 

\bigskip

3) In the \citet{2021ApJ...908....7S} modeling of the low-luminosity NGC 188 BSS WOCS 5379, regular RLOF MT occurs at the donor's early RGB phase and initially is briefly unstable. In modeling that case, a 22\% RLOF efficiency makes a significant amount of material escape from the system. The specific angular momentum of the system increases rapidly and the system expands, which then stabilizes the MT. For WOCS 4540, stronger stellar winds in the late RGB and AGB phases after mass reversal expand the system, which allows the regular RLOF efficiency to be as high as 75\% and still be stable.

\bigskip

While stable, the MT rate via regular RLOF overflow during the largest thermal pulse is still very high - $\dot{M}\sim 10^{-3}\dot{M}_{\odot}/{\rm yr}$. The MESA simulation finds that during this pulse the system goes through a very short luminosity burst. This is a difficult scenario to model in a 1-D stellar accretion simulation, and merits detailed modeling.

\bigskip

\acknowledgements

MS thanks E. Sterl Phinney and Steve Lubow for insightful comments on binary evolution. MS also thanks Robert Farmer for helping with MESA code editing. MS and RDM thank Andrew Nine, Don Dixon, Evan Linck, Marina Kounkel and Nathan Leigh for the useful suggestions in paper editing and organizing. RDM and MS acknowledge funding support from NSF AST-1714506 and the Wisconsin Alumni Research Foundation. MS also acknowledges funding support from NSF ACI-1663696 and AST-1716436. MS thanks the KITP 2022 binary program which is supported in part by the NSF PHY-1748958 during the writing of this paper, which allows MS to discuss with other visiting scholars the related physics in this paper. RDM acknowledges gratefully generous support as a Fulbright Scholar to the Universidad de Concepcion.

\bibliography{BSS_paper}
\end{CJK*}
\end{document}